Check for updates

*CORRESPONDENCE
Terry-Ann Suer,
tsuer@lle.rochester.edu

# The distribution of volatile elements during rocky planet formation

Terry-Ann Suer[1,2]*, Colin Jackson[3], Damanveer S. Grewal[4], Celia Dalou[5] and Tim Lichtenberg[6]

[1]Laboratory for Laser Energetics, University of Rochester, Rochester, NY, United States, [2]Steward Observatory, University of Arizona, Tucson, AZ, United States, [3]Earth and Environmental Sciences, Tulane University, New Orleans, LA, United States, [4]Division of Geological and Planetary Sciences, California Institute of Technology, Pasadena, CA, United States, [5]Université de Lorraine, CNRS, Centre de Recherches Pétrographiques et Géochimiques, Nancy, France, [6]Kapteyn Astronomical Institute, University of Groningen, Groningen, Netherlands

Core segregation and atmosphere formation are two of the major processes that redistribute the volatile elements—hydrogen (H), carbon (C), nitrogen (N), and sulfur (S)—in and around rocky planets during their formation. The volatile elements by definition accumulate in gaseous reservoirs and form atmospheres. However, under conditions of early planet formation, these elements can also behave as siderophiles (i.e., iron-loving) and become concentrated in core-forming metals. Current models of core formation suggest that metal-silicate reactions occurred over a wide pressure, temperature, and compositional space to ultimately impose the chemistries of the cores and silicate portions of rocky planets. Additionally, the solubilities of volatile elements in magmas determine their transfer between the planetary interiors and atmospheres, which has recently come into sharper focus in the context of highly irradiated, potentially molten exoplanets. Recently, there has been a significant push to experimentally investigate the metal-silicate and magma-gas exchange coefficients for volatile elements over a wide range of conditions relevant to rocky planet formation. Qualitatively, results from the metal-silicate partitioning studies suggest that cores of rocky planets could be major reservoirs of the volatile elements though significant amounts will remain in mantles. Results from solubility studies imply that under oxidizing conditions, most H and S are sequestered in the magma ocean, while most N is outgassed to the atmosphere, and C is nearly equally distributed between the atmosphere and the interior. Under reducing conditions, nearly all N dissolves in the magma ocean, the atmosphere becomes the dominant C reservoir, while H becomes more equally distributed between the interior and the atmosphere, and S remains dominantly in the interior. These chemical trends bear numerous implications for the chemical differentiation of rocky planets and the formation and longevity of secondary atmospheres in the early Solar System and exoplanetary systems. Further experimental and modeling efforts are required to understand the potential of chemical and physical disequilibria during core formation and magma ocean crystallization and to constrain the distributions of volatile elements in the interiors and atmospheres of rocky planets through their formation and long-term geologic evolution.

KEYWORDS

**planet formation, metal-silicate partitioning, magma-atmosphere exchange, core-mantle differentiation, volatile elements, atmosphere formation, super-earth exoplanets**

## 1 Introduction

Hydrogen, carbon, nitrogen, and sulfur are abundant volatiles in the atmospheres of terrestrial planets in the Solar System and are expected to similarly dominate the gaseous envelopes of extrasolar rocky planets (Wordsworth and Kreidberg, 2022). Depending on the prevailing thermodynamic and compositional properties, these elements can be present in atmospheres in oxidized (e.g., $H_2O$, $SO_2$, $CO_2$) or reduced forms (e.g., $H_2$, $H_2S$, CO) (Gaillard, et al., 2022). A significant proportion of a planet's volatile elements can be locked up in deep reservoirs such as cores and mantles whose bulk compositions are largely determined during early planet formation (Rubie, et al., 2007). Atmospheric outgassing (magma-gas exchange) can occur simultaneously with core-mantle differentiation while planets are hot enough to sustain magma oceans (Elkins-Tanton, 2012; Hirschmann, 2012). Prior to the outgassing of their secondary atmospheres, large rocky planets (>0.2 $M_E$) that formed early likely accreted primary H and/or He nebular atmospheres that would have interacted with their early magma oceans (Sasaki, 1990; Zahnle, et al., 2010). Although Earth lost its primordial atmosphere (Catling, et al., 2001), studies of exoplanetary systems have revealed that some rocky planets can retain such envelopes throughout their evolution (Misener and Schlichting, 2021). Atmospheres can also be strongly modified by impact-induced shock heating and erosion (Matsui and Abe, 1986; Zahnle, et al., 1988; Davies, et al., 2020; Zahnle, et al., 2020). However, from a geoscientific perspective, the two main volatile redistribution mechanisms are core formation and magma outgassing, which we discuss here with particular emphasis on recent measurements. Volatiles retained within rocky planets' mantles can reside as various mineral or fluid components and participate in geological and biochemical processes (Dasgupta and Hirschmann, 2010; Li, et al., 2013; Mikhail and Sverjensky, 2014; Armstrong, et al., 2015; Hirschmann, 2018). Significant amounts of volatile elements, along with other light elements (e.g., O and Si), are also inferred to be present in the cores of Earth and other rocky planets such a Mars, where they lower the melting temperature and density of iron, leading to partially to fully molten cores capable of sustaining a dynamo (Badro, et al., 2014; Nimmo and Schubert, 2015; Brennan, et al., 2020; Lv and Liu, 2022).

The Earth's core and mantle inherited their elemental abundances during Earth's differentiation. Core-mantle segregation involved chemical reactions between Fe and silicates in a reduced (Fe-alloy saturated) magma ocean (Wade and Wood, 2005; Wood, et al., 2006). Geochemical evidence—the abundance of refractory siderophile elements such as Ni, Co, W and Mo in mantle peridotite—coupled with metal-silicate partitioning experiments have led to inferences about the average conditions of core formation (Ringwood, 1977). The earliest framework for understanding core-mantle differentiation was provided by single-stage models in which metals accumulated at the bottom of the magma ocean before descending into the core (Stevenson, 1981; Stevenson, 1988; Karato and Murthy, 1997; Solomatov, 2007). These models assumed that the entire core and magma ocean equilibrated at a common pressure-temperature-oxygen fugacity and composition ($P$-$T$-$fO_2$-$X$) (Jones and Drake, 1986) and led to valuable insights on how redox state and planetary size affect core and mantle chemistry (Righter and Drake, 1996; Wade and Wood, 2005; Corgne, et al., 2009). Early experimental studies of elemental partitioning during core-mantle-differentiation indicated that the mantle's refractory siderophile abundances require metal-silicate partitioning at pressures from 25 to 30 GPa, equivalent to a magma ocean depth of 800 km (Li and Agee, 2001). However, later experiments found that average conditions of 45–55 GPa (~1,500 km depth) and ≥3500 K (Bouhifd and Jephcoat, 2011; Siebert, et al., 2012; Fischer, et al., 2015) could better explain the mantle's Ni and Co abundances. Large impacts may have provided the energy for such deep magma oceans and their associated episodes of core formation (Tonks and Melosh, 1993; Davies, et al., 2020), although rapid pebble accretion (Johansen, et al., 2021) could also have led to high temperatures and mantle melting.

Recent experimental studies done on different groups of elements including highly siderophiles and volatiles have indicated that one set of $P$-$T$-$fO_2$-$X$ condition cannot account for all the relevant geochemical observables such as the bulk silicate Earth (BSE) contents in highly siderophile and volatile elements. More integrated approaches have been developed which consider the dispersal of impacting materials and degree of chemical equilibration before segregating into a core and a mantle (Rubie, et al., 2003; Rubie, et al., 2007; Deguen, et al., 2014). These multi-stage approaches can account for some geochemical observables such as the moderately siderophile elemental abundances in the mantles of smaller planetary bodies, which required metal-silicate reactions at lower pressure-temperature conditions (Righter and Drake, 1997; Steenstra, et al., 2016). Though core formation in terrestrial planets is generally thought to occur at pressures and temperatures where metal and silicate phases are largely immiscible, other large-scale processes such as the segregation of a sulfide matte may accompany or follow core segregation (Rubie, et al., 2016; Steenstra, et al., 2020). Additionally, density estimates for rocky exoplanets (Otegi, et al., 2020) and the bulk chemical abundances of disintegrated planetesimals in the atmospheres of polluted white dwarfs (Bonsor, et al., 2020; Bonsor, et al., 2022) indicate that the process of planetary differentiation could be common in planetary systems.

The failure of single-stage homogenous models to find a unique $P$-$T$-$fO_2$-$X$ condition that can explain all geochemical observables suggests not only that Earth formed in stages and experienced disequilibria, but also that accreting material might have changed over time (Wade, et al., 2012). Indeed, $N$-body simulations of planetary accretion (Chambers and Wetherill, 1998; O'Brien, et al., 2014) imply that fully formed rocky bodies experience many large impacts during their growth and that radial mixing of inner and outer Solar System materials occurs concomitantly. This heterogeneous framework allows for many episodes of core formation to occur at varying pressures, temperatures, and compositions (Rubie, et al., 2003; Albarede, 2009; Schönbächler, et al., 2010). Crucially, heterogeneous models driven by the mass delivery scenarios of $N$-body simulations tend to favor the delivery of more volatile-rich materials, ultimately sourced from the outer Solar System, later in the accretion process when the potential for extreme $P$-$T$ metal-silicate reactions is at its maximum (Halliday, 2013; O'Brien, et al., 2014; Rubie, et al., 2016). However, analyses of small rocky bodies and coupled to astrophysical and geochemical models suggest that volatile elements were present in the inner Solar System well before the later stages of planetary formation (Bar-Nun and Owen, 1998; Busemann, et al., 2006; Marty, 2012; Halliday, 2013; Alexander, 2017; McCubbin and Barnes, 2019; Grewal, et al., 2021c; Deligny, et al., 2021; Grewal, 2022; Grewal and Asimow, 2023).

Another consequence of heterogeneous accretion models is the implication that rocky planets can experience several magma ocean stages throughout their evolution (Elkins-Tanton, 2012). Magma-gas interactions during magma ocean stages lead to volatile outgassing at temperature and compositional conditions related, but not equal, to those inferred for core formation (Zahnle, et al., 2010; Hirschmann, 2016; Dasgupta and Grewal, 2019; Lichtenberg, 2022). Volatiles are, by their nature, prone to partitioning into the gas; accordingly, an increasing effort is being invested into developing models to account for the atmospheres of accreting planets as large volatile reservoirs, effectively limiting incorporation into magma oceans or partitioning into cores during planetary accretion. Factors such as oxygen fugacity and magma ocean depth and volatile content strongly affect the resulting atmosphere and residual mantle compositions (Hirschmann, 2016; Grewal, et al., 2021b; Gaillard, et al., 2021; Jackson et al., 2021). Models are now being developed to account for the distribution of volatile elements into atmospheres with varying thermodynamic and compositional properties (Salvador, et al., 2017; Nikolaou, et al., 2019; Lichtenberg, et al., 2021b; Bower, et al., 2022). Although solubility establishes the equilibrium distribution of a volatile element between a primordial atmosphere and its underlying magma ocean, increasing attention is being paid to how kinetic processes, including diffusion and convective stirring within planetary interiors, may affect the outgassing and ingassing of magma oceans (Salvador and Samuel, 2023).

Partition coefficients and solubilities are the fundamental data needed to parameterize core-mantle differentiation and magma ocean-atmosphere exchange models. These parameters have been incorporated into models that are used to understand how volatile elements are fractionated and cycled among major planetary reservoirs—i.e., cores, mantles, and atmospheres—for a range of bodies, including Earth, Mars, Venus, and differentiated planetesimals such as the parent bodies of iron meteorites (Hirschmann, 2012; Wordsworth, 2016; Gaillard, et al., 2021; Jackson et al., 2021; Li, et al., 2021; Grewal, et al., 2022a; Lichtenberg, et al., 2022; Grewal and Asimow, 2023). Although early models were tuned and benchmarked to the Earth, the parameterization made available by recent partitioning and solubility measurements allows the models to be adapted to a wide range of other planetary formation scenarios. Thousands of rocky exoplanets have been discovered by surveys such as the Kepler Mission (Batalha, 2014) and the Transiting Exoplanet Survey Satellite (Kaltenegger, et al., 2019), and atmospheres of super-Earths and sub-Neptunes are now being characterized by the James Webb Space Telescope (JWST) (Mansfield, et al., 2019; Ding and Wordsworth, 2022). In particular, in its first year of observations, JWST has demonstrated its capabilities to constrain the presence or absence of large volatile envelopes around short-period rocky exoplanets (Greene, et al., 2023; Ih, et al., 2023; Moran, et al., 2023; Zieba, et al., 2023). After reasonable observational times over the next few observational cycles, JWST will be able to constrain models of the volatile contents and redox states of individual rocky exoplanets (Kempton, et al., 2023; Piette, et al., 2023). It is therefore crucial to continue efforts to model the formation and evolution of rocky planets in these exoplanetary systems (Lichtenberg, et al., 2019; Kite, et al., 2020; Lichtenberg, et al., 2021b; Dorn and Lichtenberg, 2021; Gaillard, et al., 2021; Kite and Schaefer, 2021; Lichtenberg and Krijt, 2021; Schlichting and Young, 2022; Wolf, et al., 2022). Laboratory parameterizations of volatile partitioning and solubilities are increasingly being incorporated into the models which calls for an effort to contextualize the measurements. In this review, we aim to do just that, such that the synergy between laboratory investigations and astronomical characterizations of hot exoplanets can be maximized.

# 2 Metal-silicate partitioning

## 2.1 Parameterizing experiments

In addition to gravitational segregation, core-mantle differentiation in terrestrial planets can be thought of in terms of the chemical separation of core Fe from mantle oxides and silicates. Thus, laboratory experiments study the chemical transfer of relevant species between metal alloys and silicate/oxide mixtures. The parameterization of metal-silicate partitioning reactions and experimental results has been covered extensively in several recent works (Wood, et al., 2006; Corgne, et al., 2008a; Blanchard, et al., 2017; Huang and Badro, 2018; Chidester, et al., 2022), and only a brief overview is given here.

The equilibrium partitioning of an element between metal and silicate can be thought of as an exchange reaction between molten metal (M) and oxide:

$$\mathrm{MO}_{\frac{n}{2}} + \frac{n}{2}\,\mathrm{Fe} <=> \frac{n}{2}\,\mathrm{FeO} + \mathrm{M} \tag{1}$$

where $n$ is the valence of element M. At equilibrium, the partition coefficient $D$ is the ratio of the concentration (here in terms of molar fraction $X$, but can also be in terms of weight percent, wt. %) of an element in the metal phase to that in the silicate phase:

$$D_{\mathrm{M}} = \frac{X_{\mathrm{M}}}{X_{\mathrm{MO}_{n/2}}} \tag{2}$$

The partition coefficient is related to exchange coefficient of the reaction, $K_D$, by normalizing by the partition coefficient for Fe:

$$K_D(\mathrm{M}) = \frac{D_{\mathrm{M}}}{D_{\mathrm{Fe}}^{n/2}} \tag{3}$$

That is, $K_D$ is the ratio of the partition coefficient of element M to that of iron and is independent of the Fe concentrations during the reaction.

$K_D$ is related to the equilibrium constant $K$ of reaction (1), which can be defined in terms of $K_D$ and the activity coefficient ($\gamma$) of the metal and silicate phases (E.g., Chidester, et al., 2022):

$$\log K(\mathrm{M}) = \log K_D(\mathrm{M}) + \log\frac{\gamma_{\mathrm{M}}^{\mathrm{metal}}}{\left(\gamma_{\mathrm{Fe}}^{\mathrm{metal}}\right)^{n/2}} + \log\frac{\left(\gamma_{\mathrm{FeO}}^{\mathrm{silicate}}\right)^{n/2}}{\left(\gamma_{\mathrm{MO}_{n/2}}^{\mathrm{silicate}}\right)}$$
$$= a + \frac{b}{T} + c\frac{P}{T} \tag{4}$$

Note that the oxygen fugacity of the experiments can be expressed in terms of the Fe–FeO equilibrium: $\mathrm{Fe_{alloy}} + 1/2\ \mathrm{O_2} =$ $FeO_{silicate\ melt}$, i.e., the Iron-Wüstite (IW) redox buffer (E.g., Corgne, et al., 2008a; Campbell, et al., 2009; Chidester, et al., 2022):

$$\Delta IW = -2\log_{10}\frac{a_{Fe}}{a_{FeO}} = -2\log_{10}\frac{\gamma_{Fe}X_{Fe}}{\gamma_{FeO}X_{FeO}} \quad (5)$$

Where $a$ in this case refers to the activity.

For reaction (1), $K$ is related to the Gibbs free energy of the reaction (E.g., Chapter 11 of Ganguly, 2008):

$$\log K(M) = -\frac{\Delta G^{o}}{RT} = \frac{\Delta H^{o} - T\Delta S^{o} + P\Delta V^{o}}{RT} \quad (6)$$

where $R$ is the gas constant, $T$ temperature, $P$ pressure, and $\Delta H_{o}$, $\Delta S_{o}$, and $\Delta V_{o}$ are the changes in enthalpy, entropy, and volume of reaction (1), respectively. The natural log of the partition coefficient, log $D$(M) can then be expressed in terms of the above variables and constants:

$$\log D(M) = a + \frac{b}{T} + c\frac{P}{T} - d\frac{\Delta IW}{2} - \log\frac{\gamma_{M}^{metal}}{\left(\gamma_{Fe}^{metal}\right)^{n/2}} \quad (7)$$

The constants $a$, $b$, $c$, and $d$ are determined using multivariate least-squares linear regression of the measured $D$s, $P$s, $T$s, and compositions. The parameterization can also include activity terms for the light components in the alloy and silicate melt polymerization. The expression obtained from the regression can then be used to determine partition coefficients over a wide $P$-$T$-$fO_2$-$X$ space, although care must be taken when regressing because $a$, $b$, and $c$ may significantly change with variations in experimental conditions (Walter and Cottrell, 2013). By coupling these functions to core formation models, we can calculate the re-distribution of elements between cores and mantles during differentiation.

## 2.2 Core formation models

In the single-stage core formation model, a planet is assumed to differentiate instantaneously and the re-distribution of a chemical species between core and mantle is determined at one set of conditions. Using the formalism of (Rudge, et al., 2010), the mass of species c in the body is conserved as follows:

$$x_b = Fx_{ce} + (1 - F)x_{me} \quad (8)$$

where $x_b$ is the bulk composition, $x_{me}$ is the concentration in the mantle embryo, $x_{ce}$ is the concentration in the core of the embryo, and $F$ is the body's core mass fraction, which varies in the Solar System from 0.25 (Mars) to 0.8 (Mercury). The effective partitioning $D_x$ in the embryo is defined at one set of $P$-$T$-$X$ conditions and assumed constant in single-stage models:

$$D_x = \frac{x_{ce}}{x_{me}} \quad (9)$$

In multi-stage core formation models, partition coefficients evolve with the magma ocean depth to account for multiple or continuous differentiation events (Rubie, et al., 2003). The conservation of mass of species x into the accreting planet's core and mantle can integrated numerically from Eqs 9, 10 following from Rudge et al. (2010). Similar formalisms are also developed elsewhere (Rubie, et al., 2011):

$$\frac{d}{dt}\left((1-F)Mx_m\right) = \left[(1-F)x_{me} + kF\left(x_{ce} - D_X x_m\right)\right]\frac{dM}{dt} \quad (10)$$

$$\frac{d}{dt}\left(FMx_c\right) = \left[kFD_X x_m + (1-k)Fx_{ce}\right]\frac{dM}{dt} \quad (11)$$

where $x_m$ and $x_c$ are the concentrations of species x in the mantle and core respectively of the resultant body. M is the mass of the Earth which changes with time t. $D_c$ evolves with $P$-$T$-$fO_2$-$X$ conditions as the planet accretes. This framework can be adapted to different evolution and accretion paths, including accreting bodies of different sizes and compositions and varying degrees of metal-silicate equilibration between impactors and the magma ocean.

The first term on the right-hand side of Eq. 10 represents material from the embryo's mantle that is added to the Earth's mantle, whereas the second term represents the mantle that re-equilibrated with the impactor core. Partial equilibration due to inefficient mixing between large impactors towards the end of accretion (Rubie, et al., 2003; Dahl and Stevenson, 2010) is introduced by the parameter $k$, which is the fraction of the embryo's core that equilibrates with the planet's mantle. The expressions on the right-hand side of Eq. 11 represent the two paths that the metal can take to the core: $(1 - k)\,Fc_{ce}$ is the fraction of the embryo's core that is added directly to the planet's core whereas $kFD_xc_m$ is the fraction that equilibrates with the mantle. Alternatively, partial equilibration can be implemented using a turbulent mixing parameterization derived from fluid dynamic experiments (Deguen, et al., 2014), which is identical in terms of mass conservation to the approach of Rudge et al. (2010), but the single fraction $k$ is replaced by an efficiency factor $\xi$:

$$\boldsymbol{\xi} = \frac{\boldsymbol{k}}{\mathbf{1} + \frac{\boldsymbol{D_X}}{\boldsymbol{\Delta}}} \quad (12)$$

where $\Delta$ is a metal dilution ratio that is related to the relative densities of the equilibrating metal and silicate (Deguen et al., 2014).

## 2.3 Metal-silicate partitioning behaviors of the volatile elements

The effect of core formation on the distribution of volatile elements has primarily been investigated through metal-silicate partitioning experiments at conditions relevant to small protoplanets and terrestrial planet embryos (0.5 GPa $\le p \le$ 20 GPa and $T$ up to 2500 K) in large-volume press studies (Dasgupta, et al., 2013; Roskosz, et al., 2013; Boujibar, et al., 2014; Chi, et al., 2014; Stanley, et al., 2014; Li, et al., 2015; Li, et al., 2016a; Li, et al., 2016b; Dalou, et al., 2017; Clesi, et al., 2018; Speelmanns, et al., 2018; Tsuno, et al., 2018; Grewal, et al., 2019a; Grewal, et al., 2019b; Dalou, et al., 2019; Kuwahara, et al., 2019; Malavergne, et al., 2019; Grewal, et al., 2021a; Grewal, et al., 2021b; Fichtner, et al., 2021; Jackson et al., 2021; Kuwahara, et al., 2021; Zhang and Li, 2021; Grewal, et al., 2022a; Grewal, et al., 2022b; Shi, et al., 2022; Grewal and Asimow, 2023). More recently, experiments performed in laser-heated diamond anvil cells (DACs) have extended the $P$-$T$ range of measurements on volatile elements to the average conditions thought to be relevant to core formation in the Earth and beyond (20 GPa $\le p \le$ 108 GPa, and T up to ~5500 K) (Roskosz, et al., 2013; Suer, et al., 2017; Fischer, et al., 2020; Jackson et al., 2021; Tagawa, et al., 2021; Blanchard, et al., 2022; Chidester, et al., 2022). In the following subsections, we discuss these measurements and their major implications for the distribution of volatile elements between the cores and mantles of the Earth and other rocky planets during their formation. We also provide recommended values of metal-silicate partition coefficients for a range of planetary formation scenarios (See Table 1 and details in Supplementary Appendix SA).

**TABLE 1 Recommended values of volatile element partition coefficients ($D_x$) over a range of $P$-$T$-$fO_2$-$X$ conditions relevant to different magma ocean depth/planetary size and redox conditions. Functional forms from the literature are used to calculate these partition coefficients, see Supplementary Appendix SA (H (Malavergne, et al., 2019; Tagawa, et al., 2021), N (Grewal, et al., 2021b), C (Malavergne, et al., 2019; Fischer, et al., 2020; Blanchard, et al., 2022), S (Boujibar, et al., 2014; Suer, et al., 2017). Note that these recommended values based on specific set of conditions and different values will be obtained if conditions/variables in the functions are changed. Ds are rounded off to the nearest significant decimal. Uncertainties on the Ds are dependent on the uncertainties in the fits to the functional forms and can be as large as ± 20%.**

| | Planetesimals/shallow magma ocean | | Embryo/intermediate magma ocean | | Planet/deep magma ocean | |
|---|---|---|---|---|---|---|
| | Reduced (ΔIW ~ −4) | Oxidized (ΔIW ~ −1) | Reduced (ΔIW ~ −4) | Oxidized (ΔIW ~ −1) | Reduced (ΔIW ~ −4) | Oxidized (ΔIW ~ −1 to ~1.3) |
| $P$ (GPa) | 5 | 5 | 20 | 20 | 60 | 60 |
| $T$ (K) | 2000 | 2000 | 3,000 | 3,000 | 4,000 | 4,000 |
| $D_S$ | 3 | 60 | 30 | 183 | 10 | 173 |
| $D_C$ | 316 | 10 | 43 | 1.4 | 0.5 to 20 | 0.1 to 6 |
| $D_H$ | 0.3 | 0.07 | 60 | 1.9 | 100 | 30 |
| $D_N$ | 5 | 265 | 1 | 39 | no data | no data |

### 2.3.1 Hydrogen

Hydrogen is the most cosmochemically abundant of the volatile elements and is found in significant concentrations in a wide range of meteorites (Alexander, 2019a; Alexander, 2019b; McCubbin and Barnes, 2019; Lewis, et al., 2022; Peterson, et al., 2023). The D/H ratios of various terrestrial, meteoritic, and planetary reservoirs have been crucial in constraining the origins of hydrogen on Earth, which measurements indicate could be due to contributions from different reservoirs accreted at different times throughout the Solar System's history (Albarede, et al., 2013; Halliday, 2013). Studies of hydrogen metal-silicate partitioning behavior (and solubility, discussed in section 3.1.1) have recently become important in the context of sub-Neptune exoplanets, which are thought to have retained thick primodrial H and/or He envelopes (Fulton, et al., 2017; Lichtenberg, et al., 2021a; Dorn and Lichtenberg, 2021; Rogers and Owen, 2021; Schlichting and Young, 2022). There is also isotopic evidence that the Earth and Mars could have had an early H-rich atmosphere (Tian, et al., 2005; Sharp, 2017; Pahlevan, et al., 2022). Hydrogen could also have been ingassed into Earth's magma ocean if it was in contact with such an atmosphere, and may have ultimately partitioned into the core (Wu, et al., 2018; Young, et al., 2023). Likewise, the cores of H-rich exoplanets could also have inherited nebular H, which would have influenced their total water abundances (Kimura and Ikoma, 2022).

Though hypothesized to be a low-density core component, its tendency to diffuse and exsolve makes measuring the metal-silicate partitioning of hydrogen difficult, and data remain sparse (See compilation in Figure 1); it readily forms $FeH_x$ at pressures above 5 GPa, but decomposes at ambient conditions (Badding, et al., 1991; Okuchi, 1997). Okuchi (1997) indirectly (i.e., by estimating the volume of gas bubbles) determined $D_H$ at 7.5 GPa and 1,200°C–1,500°C in large-volume press experiments to be ~1 on average and proposed that H could be a primary light element in Earth's core. More recently, large-volume press experiments have yielded new estimates for $D_H$ up to 20 GPa, as constrained by a combination of analytical techniques including electron probe microanalysis (EPMA), nuclear microprobe resonance (NMR), and electron recoil detection analysis (ERDA) inferred from the latter experiments (Clesi, et al., 2018; Malavergne, et al., 2019). Resultant $D_H$ values from these two studies are lower than unity (lithophile behavior), but thermodynamic modeling also suggested that higher pressure conditions could make H more siderophile. It is important to note that these studies measured sample compositions after quenching, and it is possible that hydrogen diffused out of their quench products, leading to large biases on $D_H$. Carbon saturation in the metals of the latter experiments could also have contributed to lowering $D_H$.

Measurements in laser-heated DAC experiments indirectly inferred H content from the unit cell expansion with *in situ* X-ray diffraction (Tagawa, et al., 2021) and found H to be siderophile, with $D_H > 29$ at 30–60 GPa and 3,100–4600 K, in agreement with theoretical calculations (Yuan and Steinle-Neumann, 2020). When coupled with a core formation model, those values imply that up to 0.6 wt. % H could have been incorporated into the Earth's core and >0.15 wt. % into the cores of planets more massive than $0.1M_E$ if water was accreted during the main stage of accretion.

### 2.3.2 Carbon

The presence of carbon in Earth's and other planetary cores has been inferred from its affinity for iron at ambient conditions and its depletion in the BSE relative to carbonaceous chondrites (CIs) (Hirschmann, 2016). The results of metal-silicate partitioning experiments at pressures up to 15 GPa in large-volume presses (Dasgupta, et al., 2013; Chi, et al., 2014; Malavergne, et al., 2019) indicate that although carbon is a highly siderophile ($D_C$ up to $10^5$) at lower pressures conditions (~1 GPa), it will partition significantly less into Fe as pressure increases (e.g., $D_C \approx 10$ at 15 GPa) (See compilation in Figure 2). Models based on these results, suggest that the cores of smaller planetesimals could retain wt. % levels of carbon, whereas those of larger bodies would be less enriched (<1 wt. % C) while their mantles would become progressively more C-rich during accretion (Kuwahara, et al., 2021). Though $D_C$ can increase as $fO_2$ decreases from ΔIW −1 to −3 (Malavergne, et al., 2019), under highly reducing conditions, Si will partition strongly into iron, limiting C dissolution in cores of very reduced bodies. This chemical trend has led to the suggestion that the Ureilite parent body or Mercury could have become carbon-saturated, possibly leading to a graphite-rich crust (Vander Kaaden and McCubbin, 2015; Keppler and Golabek, 2019; Steenstra and van Westrenen, 2020). The presence of sulfur and nitrogen in core materials can also affect the partitioning of carbon into the Fe (Grewal, et al., 2019b; Jackson et al., 2021), and immiscible C- and S-rich layers could have been contemporaneous in the cores of some planetesimals (Corgne, et al., 2008b).

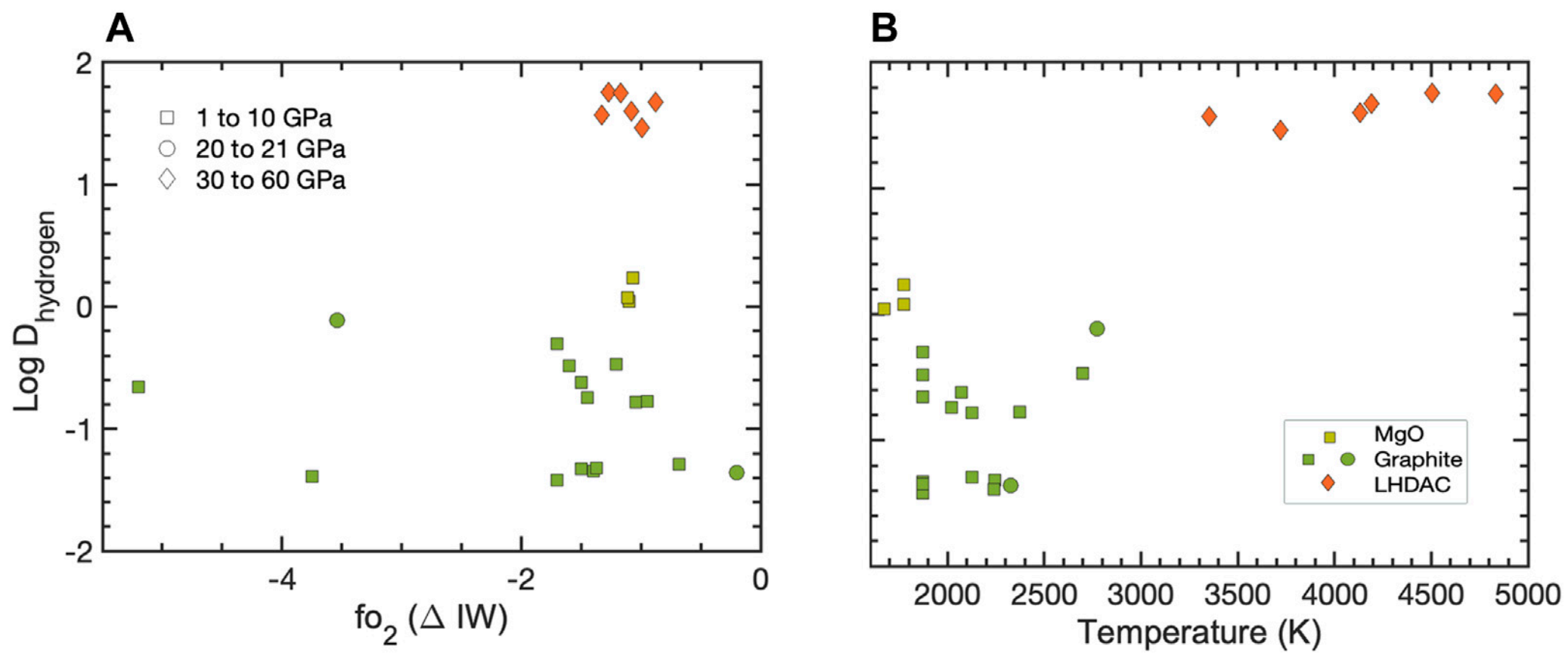

FIGURE 1
Hydrogen partition coefficients (log $D_H$) plotted as a function of oxygen fugacity, parameterized here in terms of the iron-wüstite (ΔIW) buffer **(A)** and experimental temperature **(B)**. Data sources: (Okuchi, 1997; Clesi, et al., 2018; Malavergne, et al., 2019; Tagawa, et al., 2021).

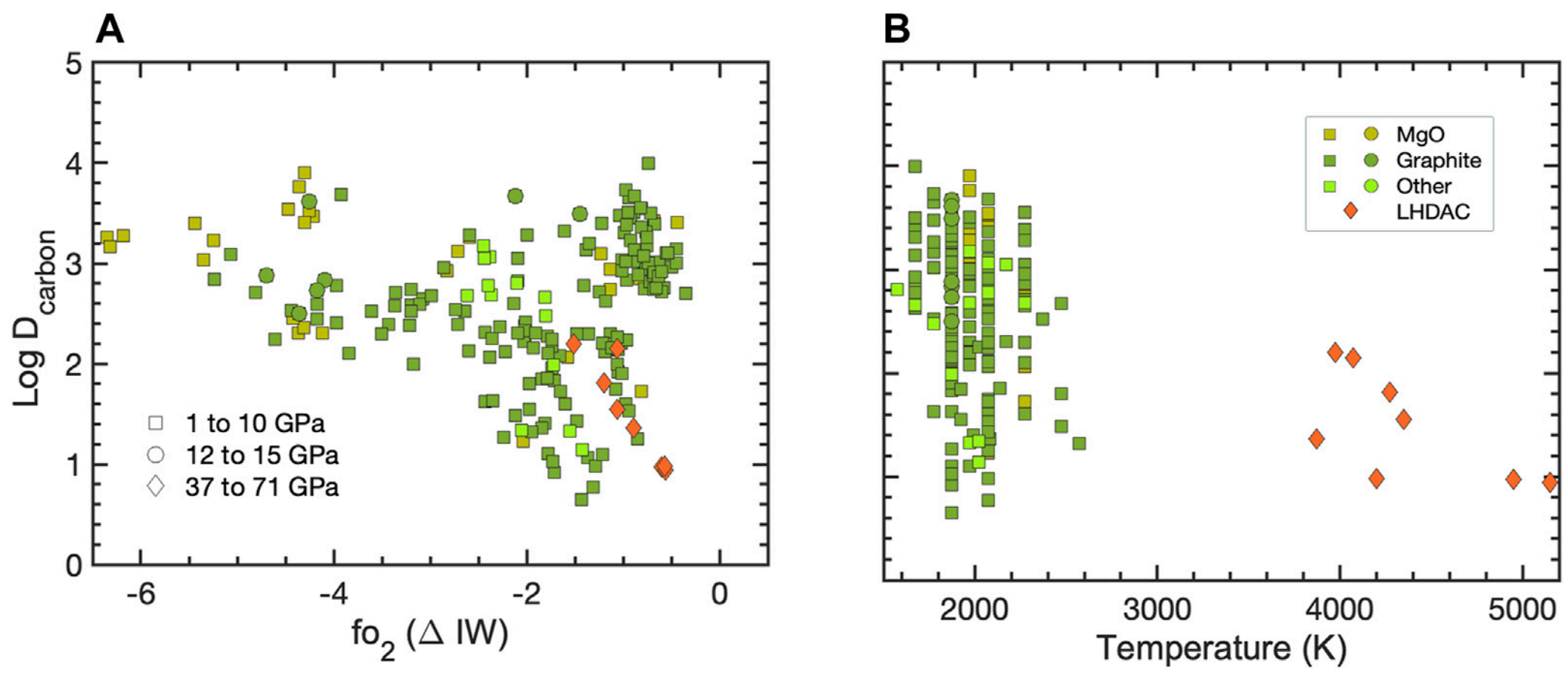

FIGURE 2
Carbon partition coefficients (log $D_C$) plotted as a function of oxygen fugacity, parameterized here in terms of the iron-wüstite (ΔIW) buffer **(A)** and experimental temperature **(B)**. Data sources: (Dasgupta, et al., 2013; Chi, et al., 2014; Armstrong, et al., 2015; Li, et al., 2016b; Dalou, et al., 2017; Tsuno, et al., 2018; Grewal, et al., 2019a; Kuwahara, et al., 2019; Malavergne, et al., 2019; Fischer, et al., 2020; Grewal, et al., 2021a; Fichtner, et al., 2021; Blanchard, et al., 2022).

Experimental results at higher *P-T* conditions in laser-heated DACs show that carbon has a lowered affinity for iron at the conditions of a deep terrestrial magma ocean (Fischer, et al., 2020; Blanchard, et al., 2022). When coupled with models of Earth's core formation and accretion, these partition coefficients imply that less than 1 wt. % C could be present in the cores of Earth and similar-sized bodies. Complementary mantle compositions could be hundreds of ppm, depending on the timing of carbon accretion.

### 2.3.3 Nitrogen

Nitrogen is the most abundant component of Earth's atmosphere and its presence in meteoritic material indicates that it was present in early planetary building blocks (Marty, 2012; Grewal, et al., 2021b; Grewal, 2022; Grewal and Asimow, 2023). Its depletion in Earth's mantle relative to C (Marty, 2012) suggests that N could have been incorporated into the Earth's core. Furthermore, it has been noted from early solubility experiments that N is increasingly incorporated into iron with increasing pressures of up to 7 GPa (Roskosz, et al., 2013; Speelmanns, et al., 2018). More recent metal-silicate partitioning experiments in large-volume presses (Dalou, et al., 2017; Grewal, et al., 2019a; Dalou, et al., 2019) found N to be siderophile ($10 < D_N < 31$) at 1–7 GPa and up to 1800°C (See Figure 3); its partitioning behavior being sensitive to P, T, $fO_2$, and the presence of light elements such as S in Fe (Grewal, et al., 2019a). The sensitivity of N partitioning to oxygen fugacity suggests that it dissolves as nitride ($N^{3-}$) in silicate melts and as neutral N in core metals (Dalou et al., 2019; Grewal et al., 2020), and this change in oxidation state makes N partitioning relatively sensitive to oxygen fugacity compared to the other volatile elements.

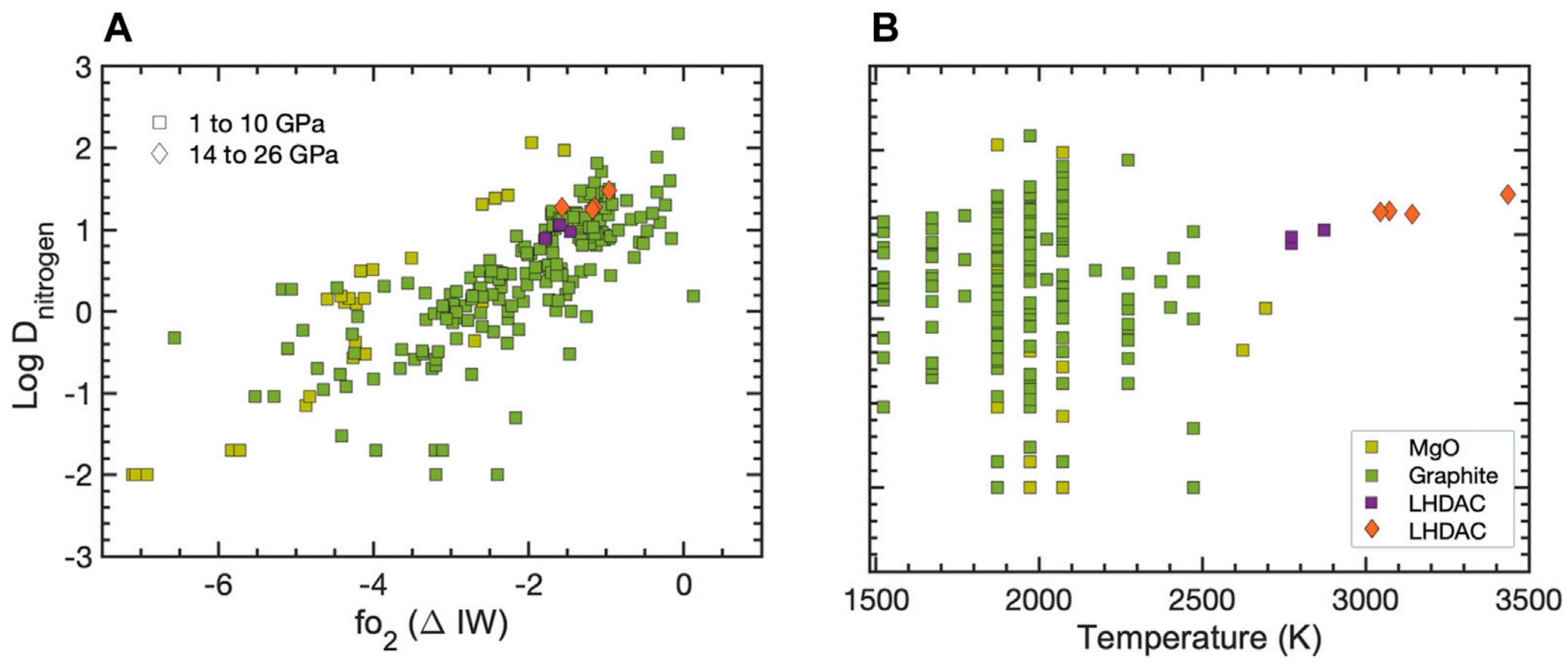

FIGURE 3
Nitrogen partition coefficients (log $D_N$) plotted as a function of oxygen fugacity, parameterized here in terms of the iron-wüstite (ΔIW) buffer **(A)** and experimental temperature **(B)**. Data sources: (Roskosz, et al., 2013; Li, et al., 2015; Dalou, et al., 2017; Speelmanns, et al., 2018; Grewal, et al., 2019a; Grewal, et al., 2019b; Dalou, et al., 2019; Grewal, et al., 2021b; Jackson et al., 2021; Shi, et al., 2022).

Nitrogen is poorly soluble in magmas at oxygen fugacities around IW (Libourel et al., 2003). Thus, magmatically active bodies can lose a large fraction of their N to their atmosphere. Combined differentiation, outgassing, and accretion models show that protoplanets that differentiated early could have been depleted in N (Grewal, et al., 2021b). However, if differentiation occurred later at the embryonic stage, nitrogen reservoirs could have been maintained within cores while the mantles would have remained depleted, a scenario which could explain the N signature of the terrestrial mantle (Grewal, et al., 2021c). Limited laser-heated DAC data demonstrate that N remains siderophile up to 26 GPa and ~3500 K (Jackson et al., 2021), further suggesting that it could be sequestered into larger planetary cores. This behavior is qualitatively different from C and S, which both become less siderophile at increased pressures and temperatures. This change in siderophile behavior among the volatile elements has significant implications for the S-C-N reservoirs on planetary embryos and larger planets (Jackson et al., 2021). The measurements compiled in Figure 3 highlight that $D_N$ increases with increasing $fO_2$. Obtaining measurements at higher *P-T* conditions will be valuable to further investigations of N incorporation into the cores of large planets.

## 2.3.4 Sulfur

Sulfur's affinity for iron at ambient conditions and its presence in iron meteorites imply that it could be a component of protoplanets and planetary cores (Dreibus and Wanke, 1985; Chabot, 2004; Gounelle and Zolensky, 2014). Combined with its cosmochemical abundance, these factors have been used to argue that S is one of the light alloying components in Earth's and other planetary cores. Its incorporation into Fe has been studied extensively in an attempt to explain the density deficit in Earths's core (i.e., relative to that of pure iron) and the properties of other planetary cores (Fei, et al., 1995; Morard, et al., 2013; Boujibar, et al., 2020; Brennan, et al., 2020). S is also known to affect the behaviors of other chemical species in various geological settings, particularly chalcophile elements (Jana and Walker, 1997; Mahan, et al., 2018) and C (Li, et al., 2016b; Tsuno, et al., 2018). Experiments in large-volume presses up to 30 GPa (Li and Agee, 2001; Boujibar, et al., 2014); showed the increasing siderophile tendency of S with increasing pressure, lending support to its incorporation into the cores of planetesimals and planetary embryos. For example, Mars, which is the size of a planetary embryo and more volatile rich than Earth, could contain up to ~20 wt. % S in its core (Brennan, et al., 2020) while it has been speculated that Mercury's core could be surrounded by an FeS layer (Namur, et al., 2016). Experiments in laser-heated DACs ~29–100 GPa, up to 5300 K (Suer, et al., 2017; Mahan, et al., 2018; Jackson et al., 2018; Chidester, et al., 2022); found $D_S$ values that are an order of magnitude lower on average than large-volume press results (ranging from less than 10–100), suggesting that sulfur's siderophile tendency does increase with pressure, but is also strongly suppressed at high temperatures (see compilation in Figure 4). When the results of laser-heated DAC experiments are included in fitting the functional form for *log* $D_S$, the entropy term is an order of magnitude larger than if the fit included only the large-volume pressure cell results. This is generally the case with laser-heated DAC studies of other elements. Redox and the chemical speciation of S have also been shown to play an important role in its partitioning behavior (Mavrogenes and O'Neill, 1999; Chidester, et al., 2022).

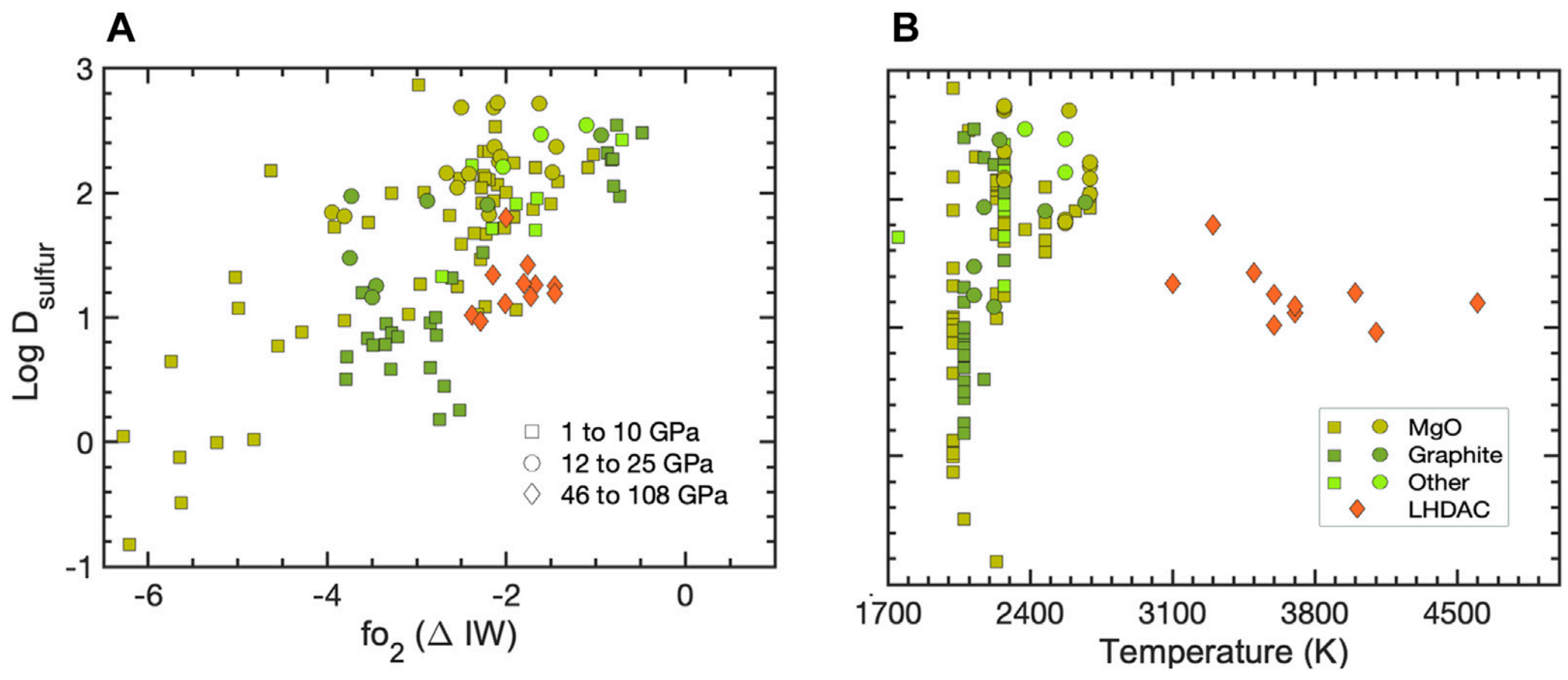

FIGURE 4
Sulfur partition coefficients (log $D_S$) plotted as a function of oxygen fugacity, parameterized here in terms of the iron-wüstite (ΔIW) buffer **(A)** and experimental temperature **(B)**. Data sources: (Li and Agee, 1996; Ohtani, et al., 1997; Wade and Wood, 2001; Chabot and Agee, 2003; Wood, 2008; Mann, et al., 2009; Rose-Weston, et al., 2009; Boujibar, et al., 2014; Suer, et al., 2017; Mahan, et al., 2018; Jackson et al., 2021; Chidester, et al., 2022).

When the functional form for $D_S$ is incorporated into models of Earth's core formation (Boujibar, et al., 2014; Suer, et al., 2017), less than 2 wt. % S can partition into the core, in agreement with cosmochemical (McDonough and Sun, 1995) and geophysical constraints (Badro, et al., 2014). However, depending on the initial bulk sulfur content and/or the timing of sulfur accretion and the efficiency of core-mantle equilibration, more sulfur could be incorporated into planetary cores. Predicted sulfur compositions of the Earth's mantle could range from a few hundred to thousands of ppm also depending on the timing of sulfur accretion (Rubie, et al., 2016; Suer, et al., 2017), overlapping with geochemical observations suggesting that the BSE contains ~200 ppm S (McDonough and Sun, 1995; Wang and Becker, 2013). It is possible that magma ocean crystallization leads to an increase in S concentrations (Rubie, et al., 2016). Above the sulfur capacity at sulfide saturation (SCSS) limit, precipitation and segregation of a sulfide-rich matte could occur on a planet-wide scale in both small and large bodies (O'Neill and Mavrogenes, 2002; Steenstra, et al., 2020), leading to a late pulse of core formation if the sulfide phase can mobilize to the core-mantle boundary.

## 2.4 Experimental limitations

In the last decade, advances in experimental and analytical techniques have enabled high-quality measurements of metal-silicate partitioning coefficients and the solubilities of volatile elements. Nonetheless, the interpretation of some of these measurements remain controversial.

Although not specific to volatile-element partitioning, ambiguity surrounds the interpretation of quench textures in both large-volume press and DAC experimental samples. Dendritic textures and overgrowths of quench rims in run products of large-volume press experiments lead to difficulty in determining stable liquid compositions. Likewise, the origin of metallic inclusions in the silicate portions of DAC run products has been controversial. Figure 5 shows examples of a large volume and laser-heated DAC experiment post quench. Large metal inclusions are usually excluded from analyses of silicate melt compositions, but smaller nanoscale inclusions (nanonuggets) are difficult to avoid, particularly with EPMA which integrates over several to tens of cubic microns. To assess this issue, NanoSIMS has been used to obtain highly resolved measurements for S and C (Suer, et al., 2017; Fischer, et al., 2020; Blanchard, et al., 2022). TEM measurements have also been performed to help resolve this controversy in recent studies (Fischer, et al., 2015; Suer, et al., 2017; Suer, et al., 2021), which suggest that nanoparticles form upon rapid quench but were dissolved during equilibrium melting. However, if nano-metal inclusions are contaminants, they could lower measured partition coefficients for the volatiles and other elements, significantly impacting the results of models that utilize these measurements.

De-volatilization and the loss of volatile species during quenching can lead to large uncertainties on experimental measurements. The degree to which a high pressure-high temperature assemblage is preserved depends on the quench rate of the experimental apparatus: example quench rates are 175°C/s in piston-cylinder experiments, ~800°C/s in multi-anvil experiments, and thousands of degrees per microsecond in DAC experiments. Magnesium-rich silicates (analogs of mafic terrestrial magma oceans) are difficult to quench to a homogenous glass, and some works have thus used basaltic and andesitic silicate compositions (Suer, et al., 2017). Quenching is particularly important for hydrogen, which is known to escape from silicate melts and metals upon recovery at ambient conditions (Okuchi, 1997). Therefore, it has been particularly difficult to obtain reliable partitioning measurements for hydrogen. Recently, the combination of EPMA, NMR and ERDA have been used successfully to measure partition coefficients for H and C (Clesi, et al., 2018; Malavergne, et al., 2019).

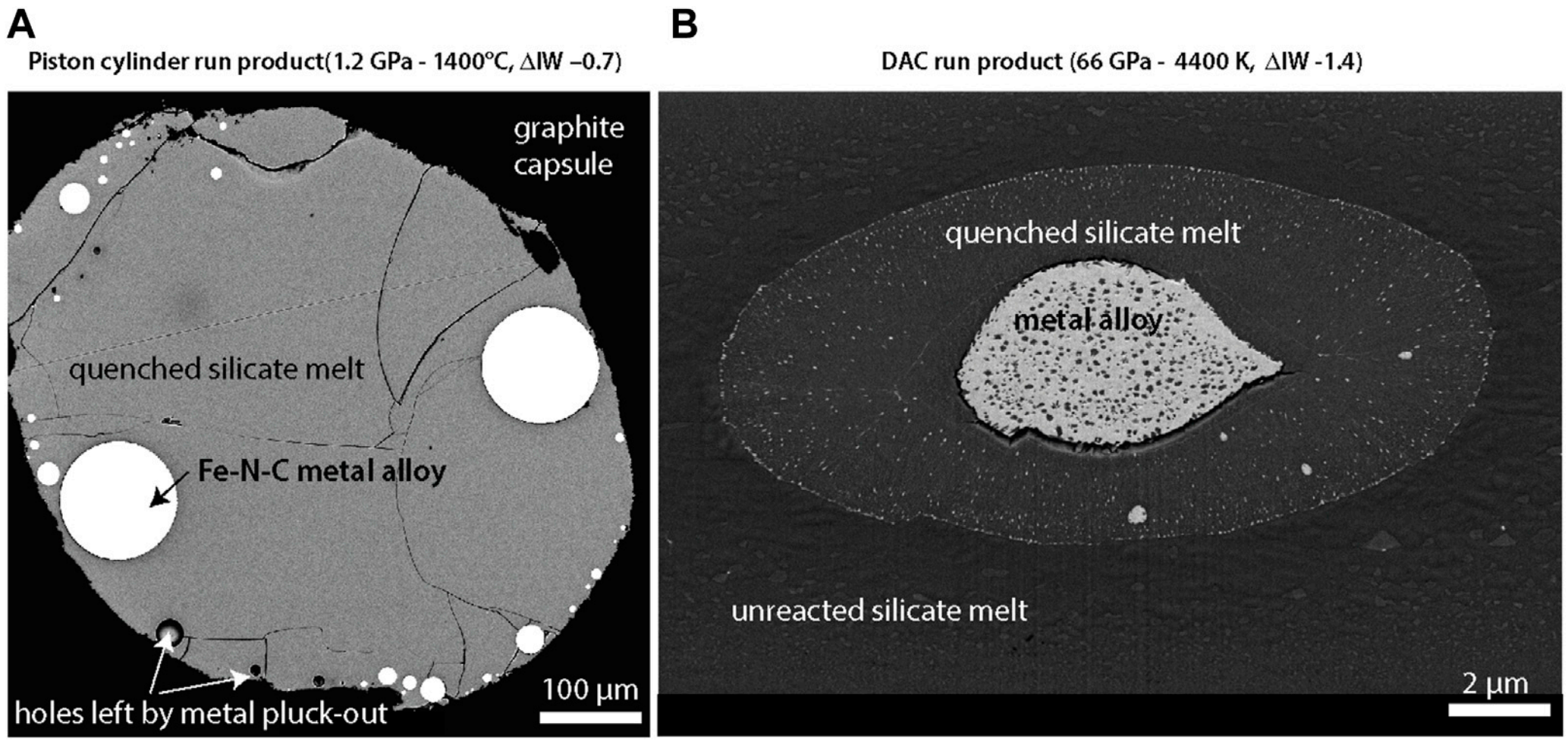

FIGURE 5
Scanning electron microscope images of quench products from metal-silicate partitioning experiments on a large-volume press **(A)** (Dalou, et al., 2017) and laser-heated diamond anvil cell **(B)**. Bright regions indicate material with higher atomic number such as Fe-Ni alloys. The quenched silicate regions of both samples contain small inclusions. In the case of diamond anvil cell experiments, inclusions can be as small as the nanometer scale.

The large concentrations of volatile elements used in the iron alloys of some experiments can also affect partition coefficients. Such concentrations are typically much higher than those observed in nature and can be in violation of Henry's law. Thus, the application of such partition coefficients in models may introduce additional, and difficult to quantify, uncertainties (Kuwahara, et al., 2019; Grewal, et al., 2021a; Grewal, et al., 2022b; Shi, et al., 2022).

Statistical regressions of measurements are used to obtain constants for the functional description of partitioning behavior. Differences have been noted between partitioning behaviors in large-volume press and DAC experiments. These have been attributed to changes in chemical behaviors in the different *P-T* regimes probed by these two types of apparatuses. This effect is perhaps best documented for S and can be seen in the different magnitudes of the regression constants obtained across different experimental techniques (Boujibar, et al., 2014; Suer, et al., 2017). Thus, a single functional form might not be sufficient to fully describe datasets spanning large *P-T* ranges. In addition, linear functional relationships might not fully describe the convolution of the IW redox buffer with the behaviors of other elements involved, and further modeling efforts are needed to deconvolve these effects. For these reasons, it is not recommended that the functional forms be extrapolated beyond the ranges of the measurements on which they are based.

# 3 Volatile element solubilities and magma-atmosphere interactions

Volatile elements distribute themselves between metal, silicate, and gas during planetary formation (Chao, et al., 2021; Lichtenberg, et al., 2022). Metal reacts with magma before its segregation to the core, whereas gas and magma continuously exchange volatiles at the atmosphere-magma ocean interface. Thus, it is expected that planetary bodies continuously redistribute their volatile element budgets as they grow and undergo new periods of core formation and mantle melting. In the previous section, our focus was on the sequestration of volatile elements within planetary cores via metal segregation in magma oceans. In this section, however, our focus is to understand how the presence of an atmosphere can affect the distribution of volatile elements during core formation, because any gaseous molecule present in an atmosphere subtracts from the budget available to metal-silicate partitioning.

## 3.1 Gas solubilities in magma oceans

Solubility is perhaps the most central parameter governing the chemical interaction of a primordial atmosphere and its underlying magma ocean. More specifically, solubility quantifies the relationship between the fugacity of a gas species and its corresponding concentration in a condensed phase. This can be expressed as:

$$\boldsymbol{X}_i = \boldsymbol{f}_i^{\alpha} \boldsymbol{S}_i \tag{13}$$

where $\boldsymbol{X}_i$ is the concentration of element $i$ dissolved in the condensed phase, $\boldsymbol{f}_i^{\alpha}$ is the fugacity of element $i$ in the gas (or fluid) raised to an exponent that relates to the stoichiometry of the dissolution reaction(s), and $\boldsymbol{S}_i$ is the solubility of element $i$ in the condensed phase. Solubility depends on the temperature and pressure of the reaction as well as the compositions of the condensed and gaseous phases. In our example, the condensed phase is the magma. Fugacity can be further deconvolved as:

$$\boldsymbol{f}_i = \gamma_i \boldsymbol{P}_i \tag{14}$$

where $\gamma_i$ and $\boldsymbol{P}_i$ are the fugacity coefficient and the partial pressure of element $i$, respectively. The high temperatures and moderate pressures in atmospheres suggest that the gases present will only have moderate deviations from ideality, and fugacity coefficients are often assumed to be unity, such that fugacity and partial pressure are equal. Eqs 13, 14 reveal the overarching role of atmospheric pressure on atmosphere-magma interactions, as it is partial pressures that drive the dissolution of volatile species into the magma.

The stability of a gas species depends on the prevailing *P-T-X* conditions within the gas phase. For example, the stability of CO with respect to $CO_2$ can be expressed as:

$$\mathbf{CO\,(g) + 0.5O_2\,(g) = CO_2\,(g)} \quad (15)$$

Le Chatelier analysis indicates that more oxidized conditions favor $CO_2$ stability compared to CO. In isolation, this reaction also predicts that $CO_2$ will be favored under lower temperature and higher pressure given that the reactants have higher entropy and volume, but all major gas species should be considered together when making specific predictions regarding how species concentration is affected by changing *P-T* conditions for any bulk gas composition.

The example above stresses the importance of oxygen fugacity in determining the stability of gas species. We note, however, that it is not precisely known how oxygen fugacity varies throughout a growing planet. It is known that, at depth, magma oceans are saturated in liquid Fe alloy and are therefore strongly reduced environments (well below the IW oxygen buffer, as inferred by the low residual FeO concentrations in the bulk silicate portions of planets), but isochemical decompression (or compression) of a magma can drive homogeneous reactions that cause auto-reduction or auto-oxidation (Zhang, et al., 2018; Armstrong, et al., 2019). Moreover, magma ocean crystallization should lead to $fO_2$ variations given the different partitioning of ferric and ferrous Fe between minerals and melts. This implies that the oxygen fugacity associated with metal-silicate reactions need not be the same as that associated with gas-magma reactions, particularly because the magma ocean may become increasingly chemically isolated form the atmosphere as solidification progresses (Bower et al., 2022). However, our goal is not to review this dynamic and the potential resulting redox stratification within planetary magma oceans (Hirschmann, 2012). Rather, we introduce this concept to stress that a wide range of oxygen fugacities are likely to impact volatile element solubilities associated with planetary formation. We correspondingly organize our review of gas solubilities below by element, and within each element by oxygen fugacity. We also emphasize that our goal is not to provide an unabridged review of volatile element solubilities in magmas, but we simply seek to highlight the basic controls on volatile element solubilities and to provide reference solubility values that may serve to accelerate the development of magma ocean-atmosphere interaction models.

### 3.1.1 Hydrogen

When reacted with oxygen, hydrogen gas forms water as (similar to Reaction 15):

$$\mathbf{H_2\,(g) + 0.5O_2\,(g) = H_2O\,(g)} \quad (16)$$

Relative to $H_2$, gaseous water molecules have a wide stability field under geological conditions, as demonstrated by the ratio of $fH_2O/fH_2$ plotted as a function of ΔIW at 2273 K (Figure 7A). The crossover of $fH_2O/fH_2$ ratio occurs near IW and is nearly independent of temperature. We note that we use thermodynamic data from the NIST WebBook to calculate equilibrium constants of reactions between gases.

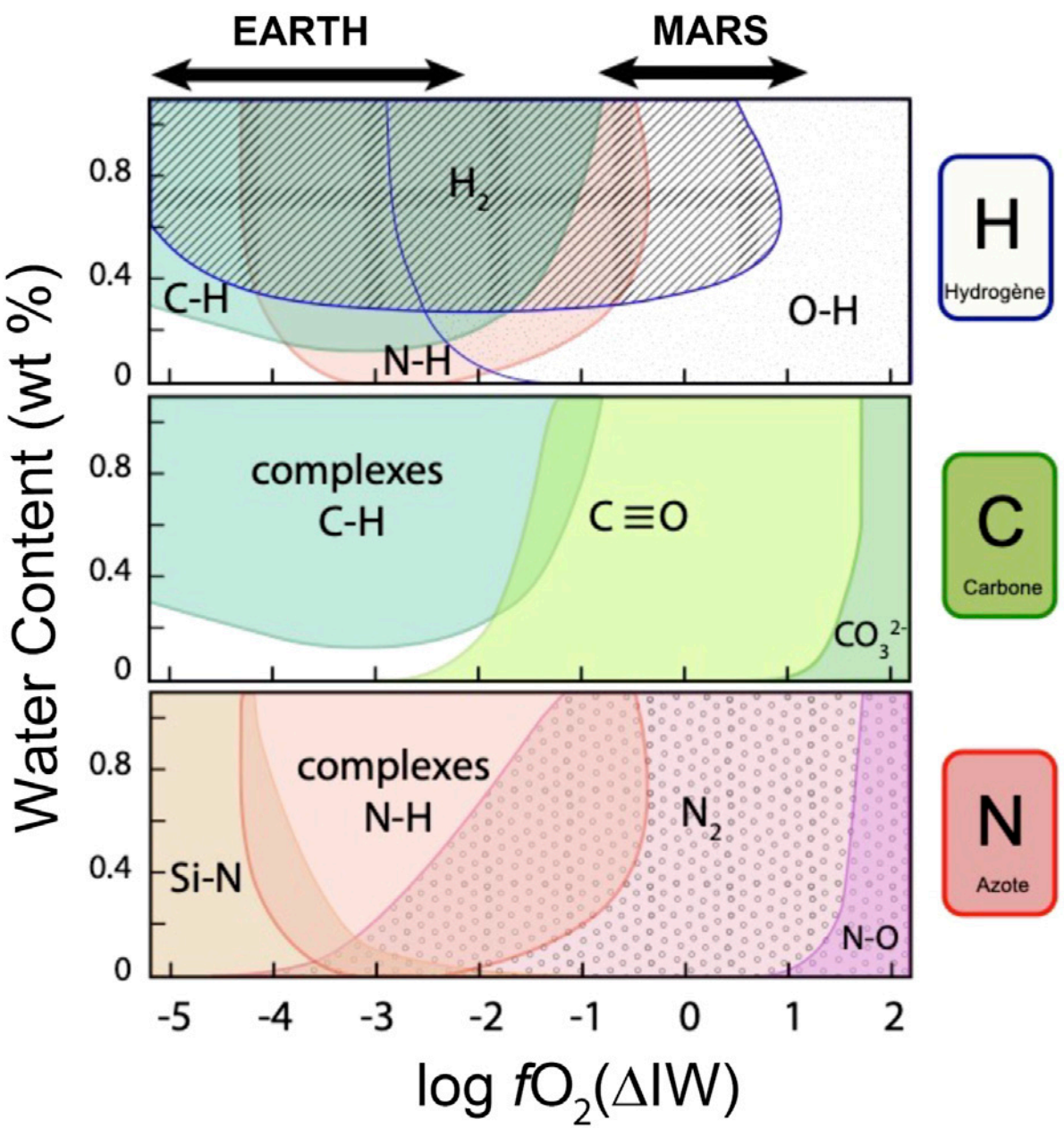

FIGURE 6
Schematic contours of H, C, and N speciation determined by Raman spectroscopy in basaltic glasses synthesized in piston-cylinder experiments at 1–2 GPa and 1,400°C–1,600°C, shown as a function of $fO_2$ (relative to the IW buffer) and H content (as equivalent $H_2O$, wt. %). Raman and FTIR measurements are from Hirschmann (2012), Armstrong et al. (2015), Dalou et al. (2019, 2022), and Grewal et al. (2020). Blank areas represent conditions at which literature data are currently lacking. This figure demonstrates that H, C, and N on their own may be present as multiple species at a given $fO_2$ or water content, and that they combine to form different molecules in natural and synthetic glasses. $fO_2$ of Earth and Mars formation are indicated. Sulfur speciation is not shown because the effect of H on S speciation has not yet been determined at $fO_2$ relevant to magma ocean conditions.

Water vapor can dissolve into magma as either OH or $H_2O$ (molecular) following the reactions:

$$\mathbf{H_2O\ (g) = H_2O\ (melt)} \quad (17)$$

$$\mathbf{H_2O\,(melt) + O^{-2}\ (melt) = 2OH^-\ (melt)} \quad (18)$$

Reactions 17 and 18 predict that low $fH_2O$ values should favor the dissolution of OH; however, with increasing OH concentration, relatively more $H_2O$ will dissolve as molecular $H_2O$. Indeed, water solubility has been experimentally shown to scale with $fH_2O^{0.5}$ at lower concentrations, but with $fH_2O$ at higher concentration. The transition from OH- to $H_2O$-dominated water solubility typically occurs at the wt. % level (Stolper, 1982), which requires $fH_2O$ > 100 bars, whereas terrestrial planets are typically estimated to contain between 100 and 1,000 ppm $H_2O$ integrated over their silicate reservoirs, atmospheres, and oceans (Peslier, 2010; Halliday, 2013). This implies that only modest $H_2O$ pressures could have been present in primordial atmospheres (~1–10 bars on average), unless large amounts of H were lost from the atmospheres (Catling, et al., 2001). Moreover, the products in Reaction 18 are favored at higher temperatures (Nowak and

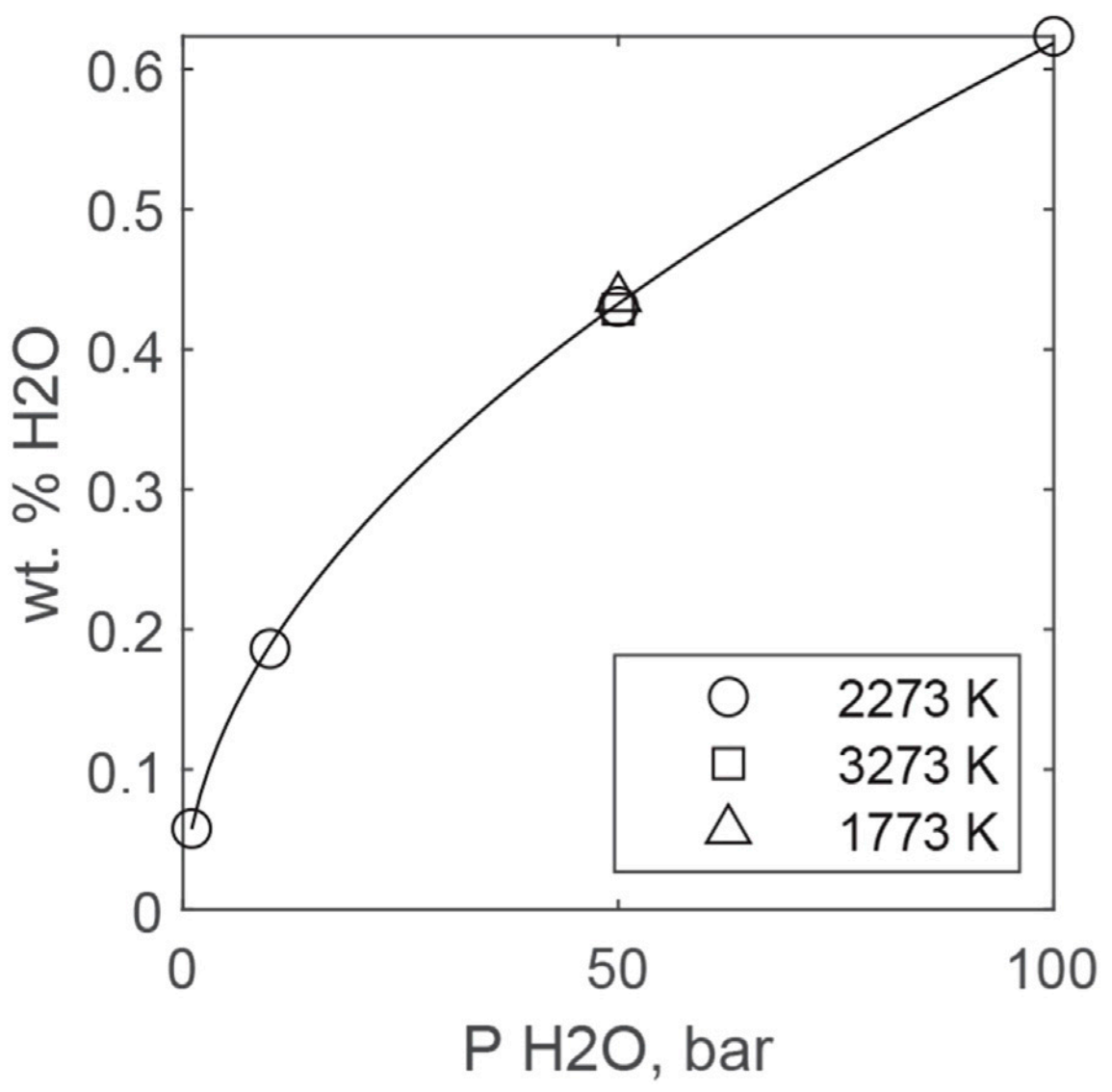

FIGURE 7
MAGMASAT predictions of $H_2O$ solubility in an ultramafic melt at different temperatures. The curve is fit to the 2273 K data and is given by Eq. 19.

Behrens, 1995), and, given that magma oceans exist at higher temperatures than modern magmatic systems, it is expected that OH is a major component of water solubility in growing planets, even under relatively reducing conditions (Figure 6).

Hydroxyl forms chemical bonds when dissolved in silicate melts, lowering melt viscosity and implying that water reacts with bridging oxygens in Reaction 18 to depolymerize the melt network. Magma ocean liquids are depolymerized, meaning that they have few bridging oxygens to support OH dissolution. Studies of water solubility in depolymerized liquids highlight that water also bonds with non-bridging oxygens, preferentially forming complexes with higher field strength cations (Mysen and Virgo, 1986; Xue and Kanzaki, 2004). It is therefore expected that water should dominantly dissolve as OH at the low $fH_2O$ conditions relevant to magma ocean liquids. A recent work suggests that peridotitic magmas contain relatively more OH than $H_2O$ compared to more polymerized magmas at equal total $H_2O$ contents, such that $H_2O$ solubility in peridotite liquids will remain dominated by OH up to 5 wt. % $H_2O$ (Bondar, et al., 2023).

Experimental determinations of water solubility in magma oceans have historically been precluded by difficulties associated with working in ultramafic systems, although recent advances in rapid-quench multi-anvil and laser levitation techniques are highly promising (Bondar, et al., 2023; Sossi, et al., 2023). In the absence of well-established data, we rely on models that allow for the *P-T-X* dependencies of water solubility determined within the realm of quenchable liquids to be extrapolated to the *P-T-X* realm of magma oceans. Several models have been developed (Newman and Lowenstern, 2002; Papale, et al., 2006; Iacovino, et al., 2021), and we use MAGMASAT (Ghiorso and Gualda, 2015) for our example because it is implemented in the popular MELTs algorithm and considers both OH and $H_2O$ dissolution.

Figure 7 plots predicted $H_2O$ solubilities from MAGMASAT for a peridotitic liquid (BSE of (McDonough and Sun, 1995) as a function of $PH_2O$ (assumed ideal) and temperature. Solubility scales close to $PH_2O^{0.5}$ and is essentially independent of temperature, although recent experimental data suggest that water solubility decreases with temperature. Fitting all predictions together yields the following equation:

$$\mathbf{H_2O_{wt.\%} = 0.0575 \cdot \mathit{PH_2O}^{0.5158}} \tag{19}$$

with $PH_2O$ in bars. The fact that MAGMASAT predicts that H2O solubility scales close to $PH_2O^{0.5}$ is consistent with OH dominating solubility up to 0.6 wt. % H2O dissolved in peridotitic magma.

Such predictions require significant extrapolation because most water solubility determinations are limited to basaltic (or move evolved) compositions and typical magmatic temperatures (<1673 K) and should therefore be treated with caution. Nevertheless, it is clear that, relative to other gases (see following subsections), water vapor remains highly soluble in ultramafic magma, even under magma ocean conditions.

Support for MAGMASAT model prediction comes from recent results from a laser levitation apparatus study (Sossi, et al., 2023). This apparatus enables quenching of ultramafic glasses reacted with controlled $fH_2O$ at temperatures that are directly applicable to magma oceans. Current data collected in systems with $fH_2O/fH_2 > 1$ yield a similar solubility relationship to that predicted above:

$$\mathbf{H_2O_{wt.\%} = 0.0605 \cdot \mathit{PH_2O}^{0.47}} \tag{20}$$

assuming a molar absorption coefficient ($\varepsilon_{3550}$) of 5.1 $m^2$/mol.

Under more reducing conditions, water vapor is progressively destabilized to produce $H_2$. Spectral evidence demonstrates that $H_2$ gas dissolves as molecular $H_2$ in magmas (Luth, et al., 1987). As a neutral, non-polar molecule, $H_2$ is thought to dissolve into the ionic porosity of the melt structure, following the well-documented behavior of noble gases (Carroll and Stolper, 1993). The radius of $H_2$ is similar to that of helium, and indeed, experiments demonstrate that $H_2$ solubility is also similar to that of helium, at least for basaltic melts (Hirschmann, 2012). The solubility of helium, or hydrogen, has not be experimentally investigated in peridotitic liquids, but their solubilities can be estimated by extrapolating the relationship of ionic porosity and solubility for helium, assuming that hydrogen solubility continues to track with helium towards more depolymerized compositions on a molar basis. To do this, we apply the algorithm developed for helium solubility in silicate melt by (Iacono-Marziano, et al., 2010) to estimate $H_2$ solubility at 1 bar and 2273 K. Taking a melt composition equal to the BSE yields the relationship:

$$\mathbf{H_{2\,wt.\%} = 4.19 \times 10^{-5} \cdot \mathit{PH_2}} \tag{21}$$

Experiments demonstrate that temperature has a modest effect on the solubilities of noble gases (Jambon, et al., 1986) although data are limited to relatively cool conditions compared to those associated with magma oceans (<1873 K). It has been suggested that higher temperatures should favor the dissolution of $H_2$ into magmas (Chachan and Stevenson, 2018; Kite, et al., 2020), and this effect could be significant given the wide temperature range over which magma oceans likely exist. The model of (Iacono-Marziano, et al., 2010) predicts an order of magnitude increase in helium solubility upon heating from 1773 to 3272 K, also suggesting the importance of high temperatures in promoting $H_2$ dissolution into magmas.

We now apply the solubility laws for H to calculate its equilibrium distribution between a magma ocean and a primordial atmosphere. Solving the equilibrium distribution for any volatile element is accomplished by solving a system of equations that include the solubility laws, statements of mass balance, and the equilibrium constants for gas phase equilibria to define the relative stabilities of gas species as a function of magma ocean-atmosphere interface *P-T-X* conditions. Mass balance statements use total mass as a constraint, but the mass balance constraint could also be the pressure of a gas species in the atmosphere or the concentration of the relevant volatile element in the magma ocean.

To connect solubility laws to mass distribution, melt concentration is scaled to mass by multiplying concentration by the mass of the assumed magma ocean. Partial pressure (fugacity) is scaled to atmospheric mass as:

$$M_x = \frac{P_x A r}{g} \tag{22}$$

where $P_x$ is the partial pressure of gas $x$ in the atmosphere, $A$ is the surface area of the planet, $g$ is the gravitational constant, and $r$ is the ratio of mass of the volatile element of interest in the gas molecule to the full mass of the gas molecule (e.g., 1 for $H_2$ and 2/18 for $H_2O$). It is clear that the distribution of a volatile element depends on the size of the planet ($A$), its gravitational field ($g$), and the depth (or mass) of the magma ocean, so these also must be specified.

We provide as an example the system of equations needed to solve the distribution of H. We apply a similar approach for other volatile elements in their respective sections.

$$K = PH_2O/\left(fO_2^{0.5}PH_2\right) \tag{23}$$

$$M_H = 4.19\times10^{-5}\cdot\frac{PH_2M_{MO}}{100} + 0.0575\cdot\frac{PH_2O^{0.5158}M_{MO}}{100} + \frac{PH_2A}{g} + \frac{PH_2OA\left(\frac{2}{18}\right)}{g} \tag{24}$$

where $K$ is the equilibrium constant for Reaction 24, $M_H$ is the total mass of H in the atmosphere-magma ocean system, and $M_{MO}$ is the mass of the magma ocean. The equilibrium constant is defined by the *P-T* conditions selected and the associated Gibbs energy change for the reaction (calculated from data tabulated on the NIST WebBook and the ideal gas law). The factors of 100 are required in the leftmost terms on the righthand side to convert from wt. % to wt. fraction. The system is solved for the partial pressures of $H_2$ and $H_2O$ ($PH_2$ and $PH_2O$). This assumes ideal gas behavior; high temperature-moderate pressure systems (hundreds of bars) only have small deviations from ideality. For example, the fugacity coefficient of steam is 0.91 at 1073 K and 500 bars (Helgeson and Kirkham, 1974) and that of $CO_2$ is 1.2 at 1500 K and 500 bars (Mel'nik, 1972), taking the standard state to be 1 bar. Higher temperatures serve only to force fugacity coefficients closer to unity. $CO_2$-$H_2O$ mixtures do have excess energies of mixing, but again, this effect is generally small at the high temperatures and moderate pressures of volatile-dominated atmospheres overlying magma oceans. The solution may require iteration between the solved pressure of the atmosphere and that assumed to calculate $K$ until there is agreement.

We note that this approach can also be extended to predict core chemistry by adding a term that includes a partition coefficient (with implied *P-T-X* conditions for metal-silicate chemical exchange) along with the available masses of silicate and metal to be reacted to define the mass of the volatile element sequestered in the core for the specified *P-T-X* conditions.

Figure 8 plots the distributions of H species between atmospheres and magma oceans as a function of $fO_2$ (relative to the IW oxygen buffer at 1 bar) and the total abundance of H. The temperature dependance of $H_2$ and $H_2O$ stability is largely canceled by referencing $fO_2$ to the IW buffer. For simplicity, we assume a total atmospheric pressure of 100 bars and a temperature of 2773 K. An assumed constant total pressure is supported by the relatively constant and high pressure that C species provide given the abundance of C in the BSE, as will be demonstrated in the next subsection.

The crossover for equal partial pressures of $H_2$ and $H_2O$ occurs near ΔIW−1 at 100 bars and 2273 K (Figure 8A). Water (steam) is much more soluble than $H_2$, and this shifts the crossover point for equal $H_2O$ and $H_2$ concentrations in magma oceans to extremely reducing conditions (<ΔIW−6), such that $H_2O$ may dominate H dissolution in magma oceans over a wide range of accretion scenarios despite the higher partial pressures of $H_2$ in atmospheres (Figure 8B). In all scenarios, the magma ocean contains more H by mass than the atmosphere (Figure 8C). H preferentially partitions into the magma ocean under oxidizing conditions, but as $fO_2$ drops and $fH_2$ rises, the distribution becomes nearly equal, indicating the increased volatility of H in reduced systems. The small masses of H in primordial atmospheres indicate that they have only a small capacity to limit the incorporation of H into cores and determine the H abundances of the bulk silicate reservoirs (i.e., materials later derived from the magma ocean-atmosphere system, that includes mantle, crust, atmosphere, and oceans).

### 3.1.2 Carbon

Under high $fO_2$, $CO_2$ is expected to be the dominant C-bearing gas species (Reaction 15). Spectroscopic results have identified that C is present as carbonate groups in magmas when reacted with $CO_2$ gas (Mysen and Boettcher, 1975; Fine and Stolper, 1986), and this observation suggests that the following set of reactions control $CO_2$ solubility in magmas:

$$CO_2\,(g) = CO_2\,(melt) \tag{25}$$

$$CO_2\,(melt) + O^{-2}\,(melt) = CO_3^{-2}\,(melt) \tag{26}$$

From these reactions, C solubility should linearly scale with $CO_2$ fugacity ($\alpha = 1$ in Eq. 13), at least up to moderate pressures, and indeed this has been experimentally documented (Dixon, et al., 1995).

Carbon solubility increases with increasing alkali and alkali Earth metal contents in magmas, which is consistent with the stability of carbonate mineral-like complexes in magmas (Lesne, et al., 2011; Duncan, et al., 2017). Higher temperatures favor the dissolution of $CO_2$ in the melt as carbonate (Konschak and Keppler, 2014). Relatively large amounts of $CO_2$ (molecular) can also dissolve into more evolved (higher $SiO_2$) melt compositions, following Reaction 25. Higher temperatures also favor reactants in Reaction 25 (Konschak and Keppler, 2014), and it is therefore possible that molecular $CO_2$ contributes to $CO_2$ solubility within magma oceans.

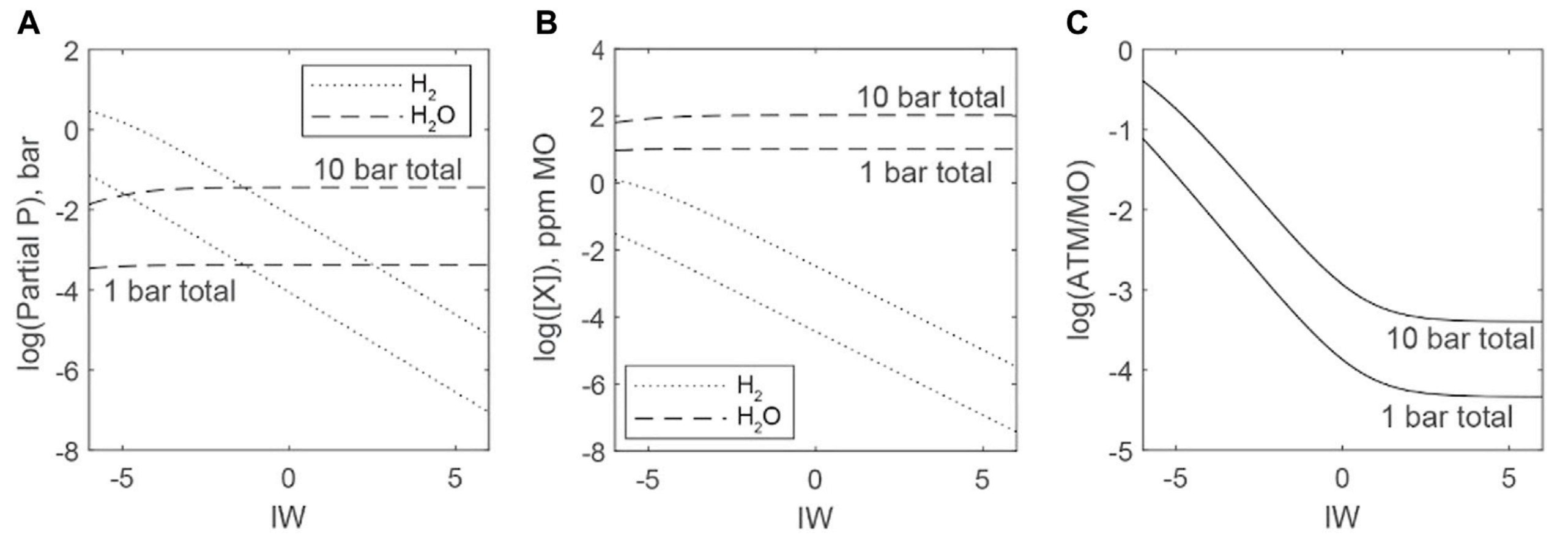

FIGURE 8
Distributions of H between atmosphere and magma ocean as functions of $fO_2$ (ΔIW). Low-H (H equal to a 1 bar atmosphere of $H_2$ or $H_2O$) and high-H scenarios (H equal to a 10 bar atmosphere of $H_2$ or $H_2O$) are considered. Masses of H in the low-H and high-H scenarios are broadly similar to those inferred for the bulk silicate reservoirs of terrestrial bodies within the solar system. All scenarios assume an atmosphere of 100 bars total pressure, a temperature of 2773 K, an Earth-sized planet, and a full mantle magma ocean. **(A)** Predictions of partial pressures of $H_2$ and $H_2O$ gases in equilibrium with a magma ocean. $PH_2O$ is nearly constant across the range of $fO_2$ considered, whereas $PH_2$ increases under more reducing conditions. **(B)** Predictions of H species concentrations for magma oceans in equilibrium with partial pressures of gases in **(A)**. **(C)** Mass ratios of H in the atmosphere versus H dissolved in magma oceans. Oxidizing conditions force $H_2O$ stability and nearly all H to dissolve into magma oceans. Reducing conditions enable a more equal distribution of H between atmosphere and magma ocean.

Sufficient experimental data have been collected to permit the development of models for $CO_2$ solubility in magmas (Dixon and Stolper, 1995; Papale, et al., 2006; Iacono-Marziano, et al., 2012). We again apply MAGMASAT to predict $CO_2$ solubility in peridotitic liquids over magma ocean *P-T* conditions (Gualda and Ghiorso, 2015) (Figure 9). Solubility scales with $PCO_2$, as suggested by Reactions 25 and 26 MAGMASAT predicts that higher temperatures are predicted to decrease solubility, but the effect is relatively minor, with solubility decreasing by a factor of 2 between 1,500 and 3000 K. Experimental work on mafic systems also supports a relatively small effect of temperature; although there is evidence that $CO_2$ solubility increases with increasing temperature, this is opposite to the predictions of MAGMASAT. Least squares fitting of MAGMASAT predictions of $CO_2$ solubility at 2273 K for 1, 10, 50, and 100 bars yields the following relationship for a BSE magma composition:

$$\mathbf{CO_{2\,wt.\%} = 2.14 \times 10^{-4} \cdot \mathit{P}CO_2} \tag{27}$$

Under more reducing conditions, $fCO_2$ will decrease while $fCO$ increases for any given temperature and pressure. With decreasing $fCO_2$, the carbonate and $CO_2$ contents of a magma will correspondingly decrease, while C species associated with the dissolution of CO will increase.

The exact nature of C species dissolved in magmas in equilibrium with CO remains relatively uncertain. Higher than expected C solubilities were observed in reduced, lunar analog melts rich in Fe, leading to the suggestion that Fe-carbonyl groups could be a major component of CO solubility (Wetzel, et al., 2013). Other spectroscopic analyses of melts, with or without Fe, reacted with variable $fCO$ have not found additional evidence that Fe-carbonyl is a significant component of C solubility. Rather, there is building evidence that single CO groups are the dominant species (Armstrong, et al., 2015; Yoshioka, et al., 2019). These groups may be molecular CO or CO species complexed to other components in the melt. Without a more detailed understanding for how CO interacts with magma, extrapolation of CO solubility determinations to magma ocean conditions will introduce some uncertainty. Nonetheless, it is clear that C becomes less soluble in magma as $fCO_2$ falls and $fCO$ rises. Yoshioka et al. (2019) developed the following equation to predict the solubility of C (wt. %) in MORB magmas for a given $fCO$:

$$\log(C_{wt.\%}) = -5.20 + 0.80\log(fCO) \tag{28}$$

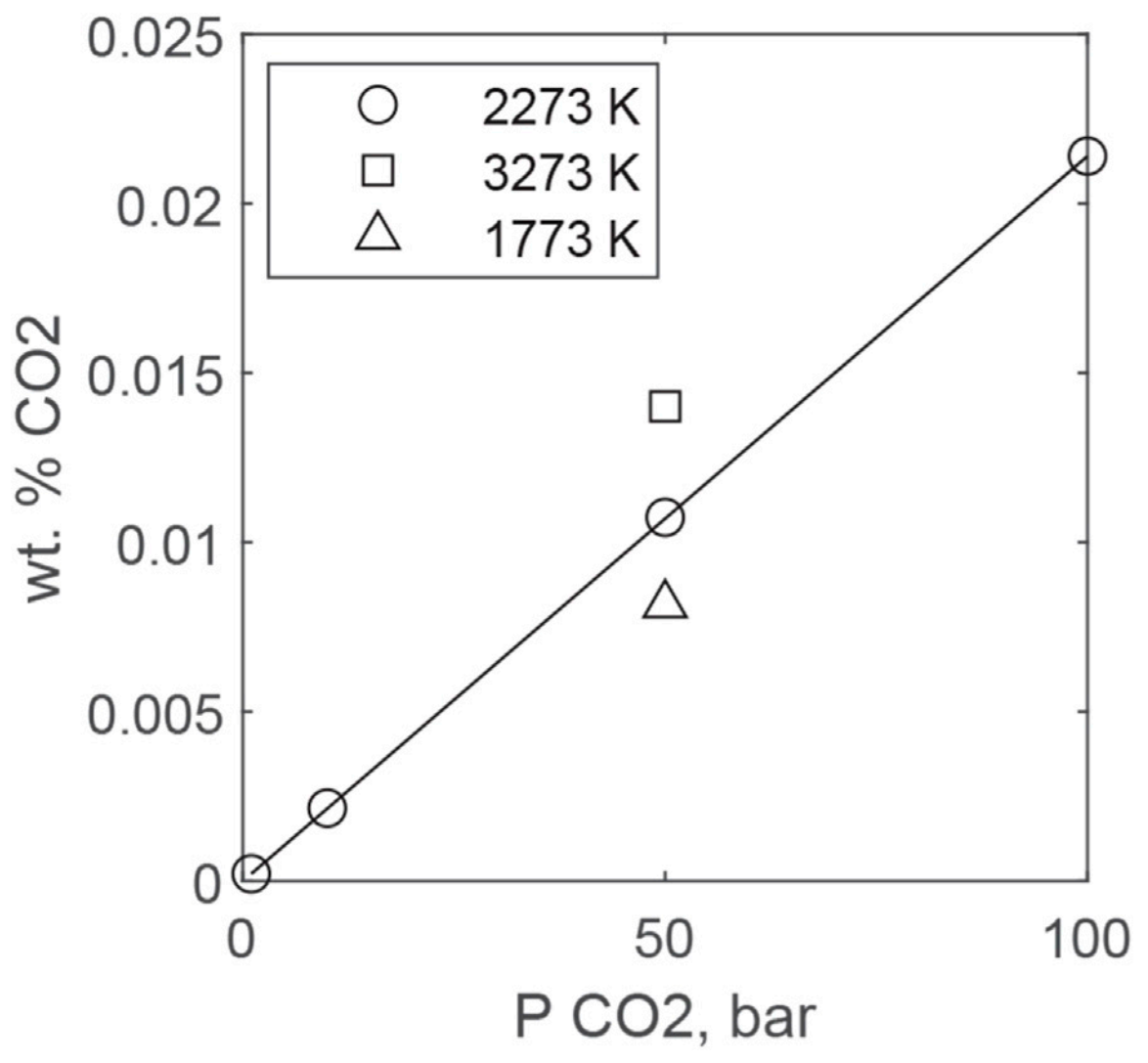

FIGURE 9
MAGMASAT predictions of $CO_2$ solubility in an ultramafic melt as a function of $PCO_2$ at different temperatures. The line is fit to the 2273 K data and is given by Eq. 27.

The unit for $fCO$ is bar. Note that the solubility of CO is expressed as reduced carbon, not CO, so statements of mass balance that incorporate Eq. 28 must take this into account.

The distribution of volatile carbon species between atmospheres and magma oceans are not predicted to be strongly dependent on the amount of carbon present in the combined system as the solubilities of all relevant species scale with their respective fugacities, with exponents close to unity.

Further decreases in $fO_2$ will progressively destabilize CO, possibly resulting in graphite/diamond/carbide saturation if overall C contents are sufficiently high. Decreases in $fO_2$ also potentially help to stabilize methane or other CH-species if a source of H is also present (Figure 6):

$$2H_2(g) + CO(g) = CH_4(g) + 0.5O_2(g) \tag{29}$$

and $CH_4$ can then dissolve into the magma following:

$$CH_4(g) = CH_4(melt) \tag{30}$$

Once dissolved in a magma, methane may undergo further reactions to produce other hydrocarbon species. Spectroscopic observations identify $CH_4$, or possibly methyl groups, dissolved in reduced, H-bearing silicate melts, supporting the importance of Reaction 29 (Mysen and Yamashita, 2010). Methane presumably dissolves into the ionic porosity of magma, following the example of $H_2$ above, but how melt-reactive hydrocarbon species interact with the silicate network has not been well explored, and their importance to magma oceans is therefore not as well established as it is for other species containing C or H.

Figure 10 plots the distribution of C species between atmospheres and magma oceans as a function of $fO_2$ and the total abundance of C. Only solubility related to $CO_2$ and CO is considered. We again assume a temperature of 2273 K, but total pressure is now determined by the combined pressure of CO and $CO_2$. Atmospheric pressure therefore decreases as CO is converted to $CO_2$ with increasing $fO_2$. We explore a high-C scenario (mass equivalent of 500 bars of C species in the atmosphere-magma ocean system) and a low-C scenario (mass equivalent 50 bars of C species in the atmosphere-magma ocean system). For context, present-day Venus has an atmosphere that is nearly 100 bars of $CO_2$. The crossover for partial pressures of CO and $CO_2$ is near IW, but higher pressures stabilize $CO_2$ gas (Eq. 28) and shift the crossover to more reducing conditions (Figure 10A). The crossover for the melt concentration of C related to CO and $CO_2$ dissolution occurs below IW because $CO_2$ species are more soluble than those related to CO (Figure 10B). Magma oceans are the larger C reservoir when $CO_2$ dominates solubility, whereas the atmosphere is the larger reservoir when CO dominates solubility (Figure 10C). The ability of the atmosphere to dominate the C budget therefore appears restricted to conditions more reducing than IW.

The discussion above for carbon assumes that the magma ocean remains undersaturated with respect to solid forms of C. High $fCO$ (and $fCO_2$) and reducing conditions promote graphite or diamond saturation following the general reaction:

$$CO(g) = 0.5O_2(melt) + C(crystal) \tag{31}$$

Once stabilized, a buffering phase such as diamond or graphite is a new volatile reservoir that must be accounted for. The physical stability of the buffering phase must also be considered. For example, graphite is expected to be buoyant in magma ocean liquids and could therefore rise to the surface (Keppler and Golabek, 2019), whereas volatile elements dissolved in alloys or stabilized in intermetallic compounds (carbide, nitrides, sulfides) may be dense and could therefore sink within a magma ocean to ultimately join the core.

### 3.1.3 Nitrogen

$N_2$ is the dominant N-bearing species in the gas phase above magma oceans (Boulliung, et al., 2020) independent of $fO_2$ (Dalou, et al., 2022; Sossi, et al., 2023). However, depending on $fO_2$, the speciation of N in silicate melts simulating magma ocean conditions falls into two domains. Under oxidized conditions, N dissolves as molecular $N_2$ with a valence state of 0 and its solubility is primarily dependent on pressure following Henry's Law (Libourel, et al., 2003; Dalou, et al., 2019; Grewal, et al., 2020; Bernadou, et al., 2021). At reduced conditions, N dissolves as ions with a −3 valence state (Libourel et al., 2003; Dalou et al., 2019; Mosenfelder et al., 2019; Boulliung et al., 2020; Grewal et al., 2020; Bernadou et al., 2021). Depending on $fH_2$, these ions could either be anhydrous nitrides ($N^{3-}$) or hydrous amines ($NH^{2-}$) and ammonia ($NH_3$) molecules (Mosenfelder, et al., 2019; Grewal, et al., 2020). $N_2$ and $NH_3$ molecules physically dissolve into the ionic porosity of silicate melts whereas ionic nitrides and amides chemically dissolve into the silicate melt network by displacing $O^{2-}$ from the silicate melt network (Libourel et al., 2003). Whereas $N_2$, $NH_3$, and $NH^{2-}$ species have been observed by Raman and FTIR (fourier transform infrared) spectroscopy in quenched silicate glasses (Dalou, et al., 2019; Mosenfelder, et al., 2019; Grewal, et al., 2020) (Figure 6), the presence of anhydrous $N^{3-}$ has only been observed by Raman by Dalou et al. (2022) at very low $fO_2$. N solubility as $N_2$ can be represented as:

$$N_2(g) = N_2(melt) \tag{32}$$

Even though hydrous N-H species are widely observed in experimental silicate melts simulating magma ocean conditions, the dissolution of N solely as $N^{3-}$ can adequately fit the observed N abundances in reduced silicate melts in both hydrous and anhydrous conditions, and can be represented as:

$$N_2(g) + 3O^{2-}(melt) = 2N^{3-}(melt) + \frac{3}{2}N_2(g) \tag{33}$$

The solubility of N as molecular $N_2$ under oxidized conditions (Eq. 31) scales with $fN_2$ and is independent of $fO_2$, whereas the solubility of N as $N^{3-}$ under reduced conditions (Eq. 32), in addition to scaling with ($fN_2$) with an exponent between 0.5 and 0.75. Although the exact $fO_2$ for the transition of N speciation depends upon several thermodynamic parameters like $T$, $P$, fluid composition, and melt composition, available evidence suggests that this transition likely occurs near the IW redox buffer for silicate

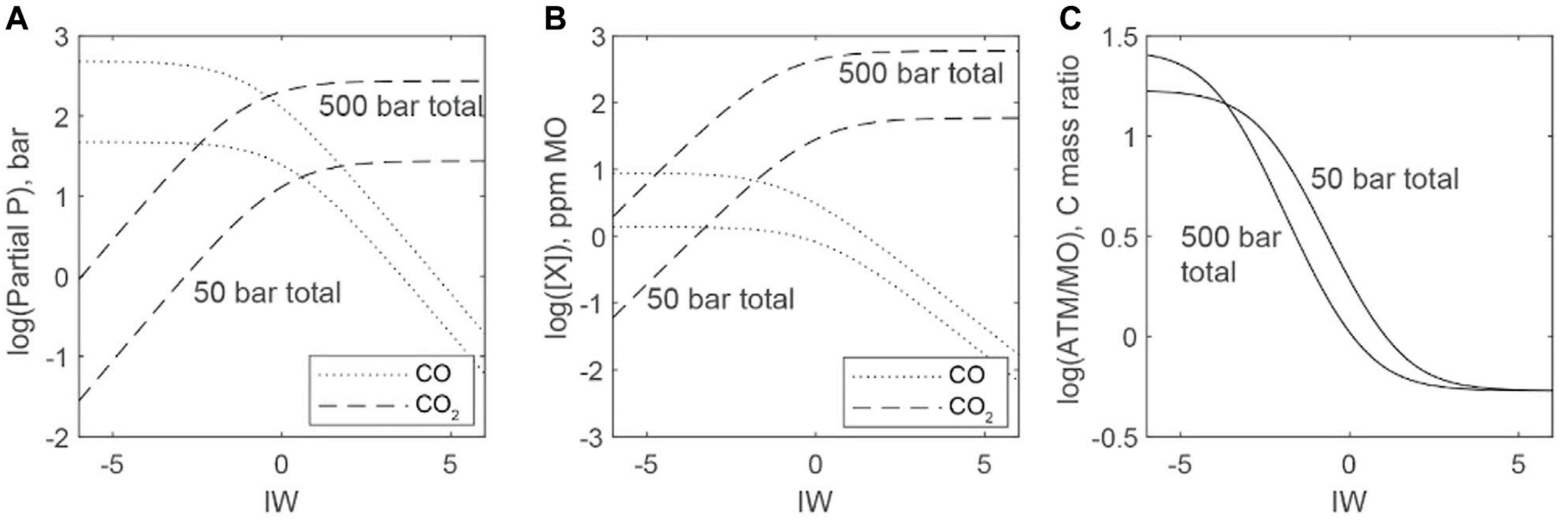

FIGURE 10
Distributions of C between atmosphere and magma ocean as functions of $fO_2$ (ΔIW). Low-C (C equal to a 50 bar atmosphere of CO or $CO_2$) and high-C scenarios (C equal to a 500 bar atmosphere of CO or $CO_2$) are considered. Masses of C in the low-C and high-C scenarios are broadly similar to those inferred for the bulk silicate reservoirs of terrestrial bodies within the solar system. All scenarios assume a temperature of 2773 K, an Earth-sized planet, and a full mantle magma ocean. Atmospheric pressure varies with the total pressure of C-bearing gas molecules. **(A)** Predictions of partial pressures of CO and $CO_2$ gases in equilibrium with a magma ocean. $PCO_2$ dominates above IW, whereas $PCO$ dominates below IW. **(B)** Predictions of C species concentrations for magma oceans in equilibrium with partial pressures of gases in **(A)**. **(C)** Mass ratios of C in the atmosphere versus C dissolved in magma oceans. Under oxidizing conditions, C is nearly equally distributed, but reducing conditions make the atmosphere the dominant C reservoir.

melts applicable to shallow magma ocean conditions (Boulliung et al., 2020; Bernadou et al., 2021).

The solubility of N in analog magma ocean silicate melts has been extensively studied by experimental studies over the last two decades for a wide range of $fO_2$, $P$, and $T$ conditions. Libourel et al. (2003) calibrated the two-species model (dissolution of N as $N_2$ and $N^{3-}$ in oxidized and reduced conditions, respectively) with experimental data at 1 bar to determine N solubility in basaltic silicate melts as a function of $P_N$ and $fO_2$:

$$\mathbf{N_{ppm} = 0.06\mathit{P}N_2 + 5.97\mathit{P}N_2^{0.5}\mathit{f}O_2^{-0.75}} \tag{34}$$

The first term on the righthand side accounts for $N_2$ solubility, and the second term accounts for $N^{3-}$ solubility. Bernadou, et al. (2021) showed that the formalism for N solubility in the silicate melt represented by Eq. 33 remains almost unaltered from 1 to 3,000 bars in the C-H-O-N system.

Figure 11 plots the distribution of N as a function of $fO_2$ between atmospheres and magma oceans. We explore a high-N scenario (mass equivalent of 10 bars of N species in the atmosphere-magma ocean system) and a low-N scenario (mass equivalent of 1 bar of N species in the atmosphere-magma ocean system). Atmospheres start to lose significant N while magma oceans start to gain significant N below IW (Figure 11A and 9 (center)). Figure 11B shows that there are essentially two regimes for N: i) above IW, nearly all N resides in atmospheres, as $N_2$ is relatively insoluble in melts; and ii) below IW −3, nearly all N resides in magma oceans, demonstrating the highly soluble nature of nitride complexes. Between these two regimes is a relatively narrow transition zone, although the $fO_2$ values within this transitional zones overlap with those inferred for magma oceans from studies of core formation. Atmospheres are the largest N reservoir when $N_2$ controls solubility to ~IW −2, but magma oceans quickly dominate the N budget as $fO_2$ drops.

### 3.1.4 Sulfur

It has long been known (Fincham and Richardson, 1954) that at $fO_2$ below the quartz-fayalite-magnetite buffer (<IW +3), S dissolves as $S^{2-}$ by displacing $O^{2-}$ from the anion sublattice. Therefore, at oxygen fugacities relevant to magma oceans (generally <IW), S solubility in the silicate melts can be represented as:

$$O^{2-}(\text{melt}) + \frac{1}{2}S_2(g) = S^{2-}(\text{melt}) + \frac{1}{2}O_2(g) \tag{35}$$

The equilibrium of this equation can be described as:

$$\ln(K_{eq}) = \ln(aS^{2-}) + \ln(fO_2^{0.5}) - \ln(fS_2^{0.5}) - \ln(aO^{2-}) \tag{36}$$

The concentration of $O^{2-}$ anions in the silicate melt greatly exceeds those of other anions, including $S^{2-}$. $O^{2-}$ concentration or activity is assumed to be constant and Eq. 36 can be modified as (assuming the activity coefficient of $S^{2-}$ to be 1):

$$\ln(S)(\text{melt}) = \ln(C_S) + \frac{1}{2}\ln\left(f\frac{S_2}{fO_2}\right) \tag{37}$$

where $C_S$, the sulfide capacity of the silicate melt, is a pseudoequilbrium constant which is controlled by the composition of the silicate melt, and primarily by its FeO content (O'Neill and Mavrogenes, 2002). Using data from previous experiments carried out at 1 atm in which a gas of known $fO_2$ (between IW −1 and IW +3) and $fS_2$ was equilibrated with silicate melts, Gaillard et al. (2022) devised an empirical relationship to determine S solubility in the silicate melts:

$$\ln(S_{ppm}) = 13.84 - \frac{26476}{T} + 0.12FeO_{wt.\%}(\text{melt}) + \frac{1}{2}\ln\left(f\frac{S_2}{fO_2}\right) \tag{38}$$

It should be noted that this equation is calibrated for FeO-rich terrestrial basalts with relatively simplified silicate melt compositions. Although new empirical models have been calibrated with experimental data for FeO-poor magmas (Namur, et al., 2016), their utility for atmosphere-magma ocean equilibration models is limited due the lack of a $fS_2$ term.

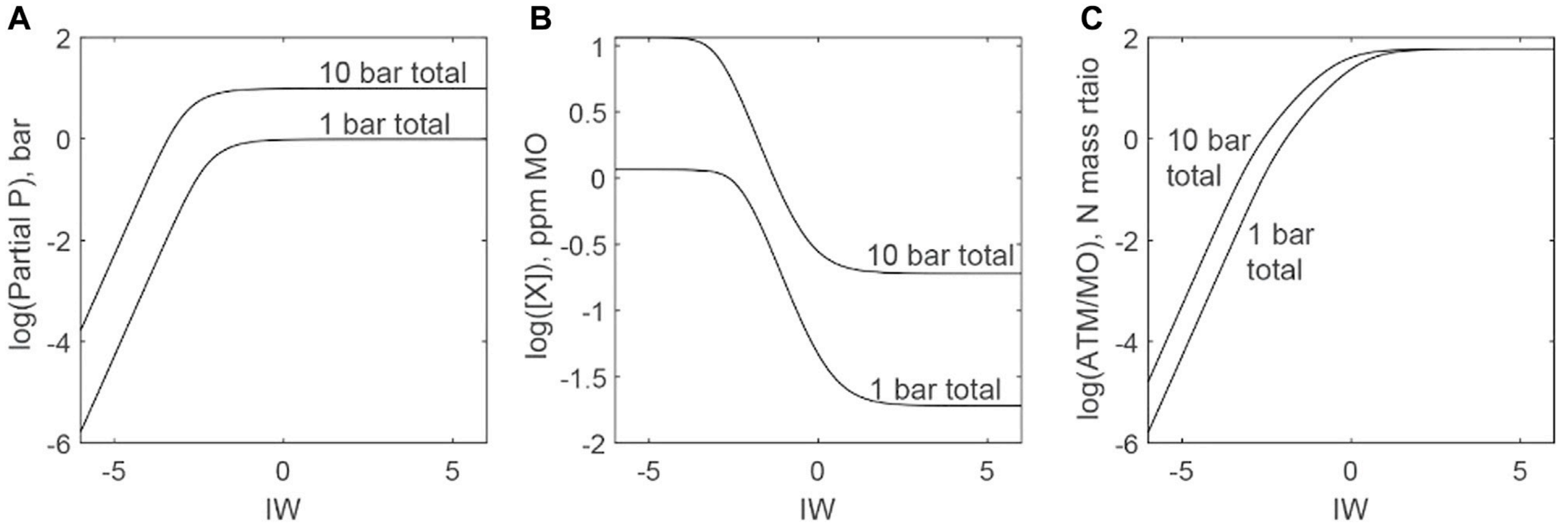

FIGURE 11
Distributions of N between atmosphere and magma ocean as functions of $fO_2$ (ΔIW). Low-N (N equal to a 1 bar atmosphere of $N_2$) and high-N scenarios (N equal to a 10 bar atmosphere of $N_2$) are considered. Masses of N in the low-N and high-N scenarios are broadly similar to those inferred for the bulk silicate reservoirs of terrestrial bodies within the solar system. All scenarios assume a temperature of 2773 K, an Earth-sized planet, and a full mantle magma ocean. **(A)** Predictions of partial pressures of $N_2$ gas in equilibrium with a magma ocean. **(B)** Predictions of N species concentrations for magma oceans in equilibrium with partial pressures of $N_2$ gas in **(C)**. **(C)** Mass ratios of N in the atmosphere versus N dissolved in magma oceans. Under oxidizing conditions, N nearly all N is present in the atmosphere, whereas reducing conditions force nearly all N to dissolve into magma oceans.

The empirical equation has FeO and $fO_2$ terms whose effects are counteract each other in Fe metal-saturated systems. Because magma ocean silicate melts are in equilibrium with metal during core-mantle differentiation, their FeO contents are thought to be lower than those of terrestrial basaltic magmas. Therefore, below IW, the effect of the $fO_2$ term dominates such that the S solubility in the silicate melt increases at increasingly reduced conditions.

Sulfur solubility is typically expressed as a function of $fS_2$ (Eq. 37), but $S_2$ is not necessarily the dominant S-bearing gas species over the *P-T-X* conditions of magma ocean-atmosphere exchange. Indeed, S is unique in its tendency to form gas molecules with other volatile elements. Hydrogen sulfide is produced by $S_2$ when exposed to elevated $fH_2$, and COS is produced by $S_2$ when exposed to elevated $f$CO. This dynamic makes predicting the atmosphere-magma ocean distribution of S correspondingly more complicated than for H, C, or N.

Figure 12 plots the stability of S-bearing species for an atmosphere with 100 bars of total pressure. The gas phase is 10 mol% H, 0.03 mol% S, and the balance is C species. Temperature is fixed at 2273 K, $fO_2$ ranges between ΔIW −4 and ΔIW +2. Hydrogen sulfide is the most abundant S-bearing gas, followed by COS, under reducing conditions, whereas $SO_2$ is most abundant under oxidizing conditions. $S_2$ remains a minority species over the entire range of $fO_2$.

Figure 12B plots the mass of S in magma oceans for our nominal scenario (10 mol% H, 0.03 mol% S, and the balance is C species), a high-S scenario (10 mol% H, 0.3 mol% S, and the balance C is species), and a high-H scenario (30 mol% H, 0.03 mol% S, and the balance C species). The mass of S in the magma ocean is calculated using Eq. 37. All atmospheres are 100 bars total, and equilibrium is calculated for 2273 K. The horizontal solid lines bracket estimated S masses for the BSE in Figure 12B. Sulfur is highly soluble in magmas, and this manifests as the much larger masses of S dissolved in magma oceans compared to the equilibrium masses of S present in atmospheres (Figure 12A).

Eq. 37 assumes that $HS^-$species do not contribute to S solubility, as the compositional dependence of S solubility does not include a term related to $fH_2$. With this assumption, the stabilization of $H_2S$ gas with increasing $fH_2$ under reducing conditions acts to make S more volatile. An example of this is provided in Figure 12 as H scenario. There is debate as to whether $HS^-$species are significant for S solubility (Baker and Moretti, 2011), so Eq. 37 may underestimate S solubility in $H_2$-rich systems. Nevertheless, it is clear that S is relatively soluble in melts and that primordial atmospheres contain only a small amount of S.

## 3.2 Beyond an equilibrium model

The analysis above reviewed solubility laws and mass balance statements to calculate the equilibrium model for the distribution of H, C, N, and S between magma oceans and atmospheres. But this framework is not complete and much work is needed to make new experimental measurements of solubility under the *P-T-X* conditions of magma oceans, including high temperatures, high pressures, ultramafic melt compositions, and highly reducing conditions. Nevertheless, applying an equilibrium model with a single *P-T-X* condition for chemical exchange within a system as large and dynamic as a combined magma ocean and atmosphere must fall short because *P-T-X* conditions vary with time and heliocentric location as planets grow.

As magma oceans cool, the interface temperature with the atmosphere will change (Lichtenberg, et al., 2021a; Lichtenberg, et al., 2022). The ability of a magma ocean and atmosphere to remain in equilibrium relies on mass exchange across their interface, but also the nature and associated kinetics of this exchange remain areas of study (Salvador, et al., 2017). The diffusion of volatiles from the interface may be important here, in addition to the convective patterns of the underlying magma ocean and its associated timescales and length scales for mixing (Solomatov, 2007). With continued cooling, magma oceans crystallize, leading to a new era of evolution in which volatiles will partition between minerals and melts, depending on the elemental exchange between the magma and its overlying atmosphere, and hence the formation of a protocrust (Bower, et al., 2022). Some minerals may be major hosts for volatiles, but many likely exclude volatiles from their structures. Crystallization should therefore lead to net outgassing. Outgassing may be facilitated by bubble nucleation, bubble growth, and the diffusion of dissolved volatiles in the magma towards bubbles (Ikoma, et al., 2018). Again, kinetics may be important in determining the chemical distribution of volatiles within the magma ocean-atmosphere system. Crystallization also offers the possibility of trapping buoyant magmas in the cumulate piles of magma oceans. Trapped magma provides a physical, rather than a chemical, mechanism for retaining volatiles in magma oceans (Hier-Majumder and Hirschmann, 2017) and is highly relevant for inhomogeneous crystallization paths in Earth- or super-Earth-sized exoplanets (Moore and Cowan, 2020). The amount of trapped liquid is also an expression of kinetic processes and the competition of melt migration timescales against crystallization timescales. Finally, recent work suggests that large impacts may result in transient, highly energetic periods in which the distinction between atmosphere and surface may vanish (Caracas and Stewart, 2023). Upon cooling, these atmospheres would shift from being silicate-rich to rich in moderately volatile elements. Ultimately, all of these processes must be accounted for to understand how much nature deviates from the equilibrium framework. In light of associated degeneracy, information from exoplanetary systems may provide additional constraints on the redox diversity and crystallization pathways of rocky planets within and beyond the Solar System.

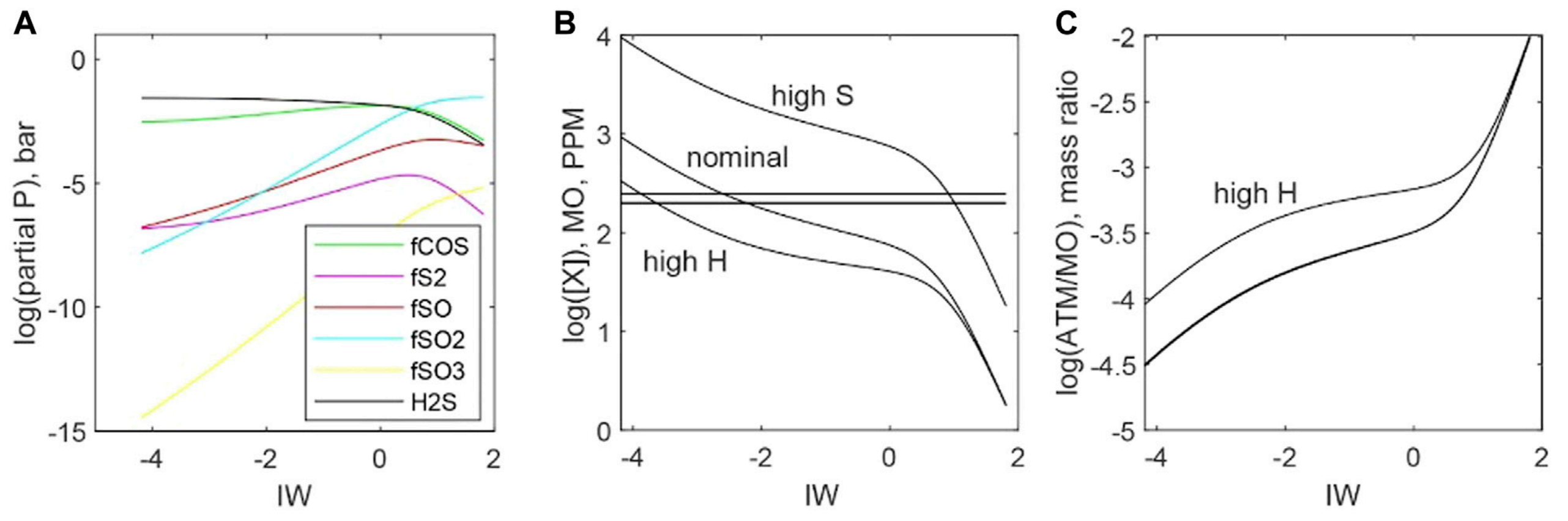

FIGURE 12
Distributions of sulfur between atmosphere and magma ocean as functions of $fO_2$ (ΔIW). **(A)** An example calculation of S gas species partial pressures for a 100-bar atmosphere with a nominal composition of 10 mol% H, 0.03 mol% S, and the balance is C species at 2273 K. **(B)** Predicted concentrations of S (dissolved in magma oceans for a nominal atmosphere, a high-S atmosphere (10 mol% H, 0.3 mol% S, and the balance is C), and a high-H atmosphere (30 mol% H, 0.03 mol% S, and the balance is C species). The horizontal lines bracket the estimated range of S concentrations in the bulk silicate Earth (McDonough and Sun, 1995) **(C)** Mass ratios of S in the atmosphere versus S dissolved in magma oceans. All scenarios predict that most S is present within magma oceans.

## 4 Summary and outlook

The cores of rocky planets could host significant amounts of volatile elements. However, a major finding from the recent metal-silicate partitioning literature is that high pressures and temperatures can have opposing effects on the partition coefficients, and on average can lower the siderophile tendencies of some elements including volatiles S and C. This could result in decreased incorporation of these elements into Fe cores during late-stage core formation in larger rocky planets. Though partial equilibration of large impactors during the giant impact phase of planetary growth could result in the transfer of unequilibrated volatile-rich materials to cores. More measurements at high P-T are needed to ascertain trends for N and H. Both N and S appear to show increased affinity for iron while C partitions less into the metal at more oxidizing conditions. These change in siderophile behaviors of the volatile elements has significant implications for the volatile reservoirs as planets grow. When coupled with accretion models, the metal-silicate partitioning measurements support heterogeneous accretion scenarios whereby volatiles were delivered towards the later stages of core formation. This is in contrast with the increasing evidence for the volatile-rich nature of inner Solar System planetesimals which could have strongly influenced the volatile budgets of terrestrial planets. Though most of the studies reviewed here were applied to the formation of the Earth and other rocky Solar System bodies, it has been inferred that core segregation likely occur during the formation of rocky planets in extrasolar planetary systems and in planets larger than the Earth such as super-earths and sub-neptunes (E.g., Otegi, et al., 2020). Calculations have indicated efficient mixing between iron and rocky materials (Wahl and Militzer, 2015) and, similarly, volatiles and rocky materials (Dorn and Lichtenberg, 2021; Kovačević, et al., 2022; Vazan, et al., 2022) would occur in such planets, and it would be beneficial to obtain measurements at higher P-T and a wider range of compositions than the current dataset to better understand metal-rock and volatile-rock differentiation mechanisms in super-Earths.

In the magma ocean stage of growth, significant amounts of volatiles can dissolve into the silicate parts of planets, in which solubilities and redox state determine the availability of volatiles to outgassing. The recent solubility studies indicate that under oxidizing conditions, most H and S are sequestered in the magma ocean, most N is outgassed to the atmosphere, and C is nearly equally distributed between the atmosphere and the interior. Under reducing conditions, nearly all N dissolves in the magma ocean, the atmosphere becomes the dominant C reservoir, H becomes more equally distributed between the interior and the atmosphere, and S remains dominantly in the interior. Coupling of measurements with microphysical to large-scale models is a direction for future work that could provide further clarity on some of these multi-scale processes. Finally, the understanding gained from these studies can guide the interpretation of new measurements of exoplanetary atmospheres, which can in turn provide additional constraints to assess the bulk compositions, formation histories, and the abilities of rocky planets to develop conditions suitable for the origin and sustainability of life as we know it.

Recent first principles/*ab initio* molecular dynamics and density functional theory (DFT) simulations have been used to gain further insights on how volatile elements are distributed among cores, magmas, and atmospheres. For example (Solomatova and Caracas, 2019; Davis, et al., 2022), found C coordination in silicate melts to be largely pressure dependent, implying that terrestrial magma oceans could have contained oxidized carbon polymers. The latter study also found strong clustering between C and Fe, which increases as a function of pressure, suggesting C could be present in cores. Molecular dynamics studies of Fe-N compounds lead to the conclusion that Earth's core likely contains very low amounts of nitrogen (Bajgain, et al., 2019), though this does not rule out higher N contents in the cores of other planets. DFT studies have been used to argue for the preferential incorporation of H in the core (Li, et al., 2020; Yuan and Steinle-Neumann, 2020). If validated by meaurements, *ab initio* methods could provide a way to estimate partition coefficients and solubilities at a wider range of conditions than are achievable with current experiments. A Bayesian inference approach might also provide a way to extrapolate these quantities to conditions beyond the measurements (e.g., Gaffney, et al., 2022).

## Author contributions

T-AS, CJ, DG, and CD contributed to the collection and curation of the data. All authors contributed to the article and approved the submitted version.

## Acknowledgments

The results reported herein benefitted from collaborations and/or information exchange within NASA's Nexus for Exoplanet System Science (NExSS) research coordination network sponsored by NASA's Science Mission Directorate and project "Alien Earths" funded under Agreement No. 80NSSC21K0593. CRMJ acknowledges support from NASA Emerging Worlds Grant 80NSSC21K0377. TL was supported by the Branco Weiss Foundation. CD acknowledges ANR grant JCJC CSI Planet. DG was funded by a Barr Foundation Postdoctoral Fellowship by California Institute of Technology. This work was supported by the AEThER project, funded by the Alfred P. Sloan Foundation under grant No. G202114194. This material is based upon work supported by the Center for Matter at Atomic Pressures (CMAP), a National Science Foundation (NSF) Physics Frontiers Center, under Award PHY2020249. The authors thank Vincent Clesi for sharing data tables for hydrogen.

## Conflict of interest

The authors declare that the research was conducted in the absence of any commercial or financial relationships that could be construed as a potential conflict of interest.

The handling editor MR declared a past collaboration with the author CD.

## Publisher's note

## Supplementary material

The Supplementary Material for this article can be found online at: https://www.frontiersin.org/articles/10.3389/feart.2023.1159412/full#supplementary-material

## References

Albarede, F. (2009). Volatile accretion history of the terrestrial planets and dynamic implications. *Nature* 461 (7268), 1227–1233. doi:10.1038/nature08477

Albarede, F., Ballhaus, C., Blichert-Toft, J., Lee, C. T., Marty, B., Moynier, F., et al. (2013). Asteroidal impacts and the origin of terrestrial and lunar volatiles. *Icarus* 222 (1), 44–52. doi:10.1016/j.icarus.2012.10.026

Alexander, C. M. D. (2017). The origin of inner Solar System water. *Philosophical Trans. R. Soc. A Math. Phys. Eng. Sci.* 375 (2094), 20150384. doi:10.1098/rsta.2015.0384

Alexander, C. M. D. (2019a). Quantitative models for the elemental and isotopic fractionations in chondrites: the carbonaceous chondrites. *Geochimica Cosmochimica Acta* 254, 277–309. doi:10.1016/j.gca.2019.02.008

Alexander, C. M. D. (2019b). Quantitative models for the elemental and isotopic fractionations in the chondrites: the non-carbonaceous chondrites. *Geochimica Cosmochimica Acta* 254, 246–276. doi:10.1016/j.gca.2019.01.026

Armstrong, K., Frost, D. J., McCammon, C. A., Rubie, D. C., and Boffa Ballaran, T. (2019). Deep magma ocean formation set the oxidation state of Earth's mantle. *Science* 365 (6456), 903–906. doi:10.1126/science.aax8376

Armstrong, L. S., Hirschmann, M. M., Stanley, B. D., Falksen, E. G., and Jacobsen, S. D. (2015). Speciation and solubility of reduced C–O–H–N volatiles in mafic melt: implications for volcanism, atmospheric evolution, and deep volatile cycles in the terrestrial planets. *Geochimica Cosmochimica Acta* 171, 283–302. doi:10.1016/j.gca.2015.07.007

Badding, J. V., Hemley, R. J., and Mao, H. K. (1991). High-pressure chemistry of hydrogen in metals: *in situ* study of iron hydride. *Science* 253 (5018), 421–424. doi:10.1126/science.253.5018.421

Badro, J., Alexander, S. C., and John, P. B. (2014). A seismologically consistent compositional model of Earth's core. *Proc. Natl. Acad. Sci.* 111 (21), 7542–7545. doi:10.1073/pnas.1316708111

Bajgain, S. K., Mookherjee, M., Dasgupta, R., Ghosh, D. B., and Karki, B. B. (2019). Nitrogen content in the Earth's outer core. *Geophys. Res. Lett.* 46 (1), 89–98. doi:10.1029/2018gl080555

Baker, D. R., and Moretti, R. (2011). Modeling the solubility of sulfur in magmas: a 50-year old geochemical challenge. *Rev. Mineralogy Geochem.* 73 (1), 167–213. doi:10.2138/rmg.2011.73.7

Bar-Nun, A., and Owen, T. (1998). Trapping of gases in water ice and consequences to comets and the atmospheres of the inner planets. *Sol. Syst. Ices* 1998, 353–366.

Batalha, N. M. (2014). Exploring exoplanet populations with NASA's Kepler Mission. *Proc. Natl. Acad. Sci.* 111 (35), 12647–12654. doi:10.1073/pnas.1304196111

Bernadou, F., Gaillard, F., Füri, E., Marrocchi, Y., and Slodczyk, A. (2021). Nitrogen solubility in basaltic silicate melt - implications for degassing processes. *Chem. Geol.* 573, 120192. doi:10.1016/j.chemgeo.2021.120192

Blanchard, I., Rubie, D., Jennings, E., Franchi, I., Zhao, X., Petitgirard, S., et al. (2022). The metal–silicate partitioning of carbon during Earth's accretion and its distribution in the early solar system. *Earth Planet. Sci. Lett.* 580, 117374. doi:10.1016/j.epsl.2022.117374

Blanchard, I., Siebert, J., Borensztajn, S., and Badro, J. (2017). The solubility of heat-producing elements in Earth's core. *Geochem. Perspect. Lett.* 5, 1–5. doi:10.7185/geochemlet.1737

Bondar, D., Withers, A. C., Whittington, A. G., Fei, H., and Katsura, T. (2023). Dissolution mechanisms of water in depolymerized silicate (peridotitic) glasses based on infrared spectroscopy. *Geochimica Cosmochimica Acta* 342, 45–61. doi:10.1016/j.gca.2022.11.029

Bonsor, A., Carter, P. J., Hollands, M., Gänsicke, B. T., Leinhardt, Z., and Harrison, J. H. D. (2020). Are exoplanetesimals differentiated? *Mon. Notices R. Astronomical Soc.* 492 (2), 2683–2697. doi:10.1093/mnras/stz3603

Bonsor, A., Lichtenberg, T., Drazkowska, J., and Buchan, A. M. (2022). Rapid formation of exoplanetesimals revealed by white dwarfs. *Nat. Astron.* 7, 39–48. doi:10.1038/s41550-022-01815-8

Bouhifd, M. A., and Jephcoat, A. P. (2011). Convergence of Ni and Co metal–silicate partition coefficients in the deep magma-ocean and coupled silicon–oxygen solubility in iron melts at high pressures. *Earth Planet. Sci. Lett.* 307 (3-4), 341–348. doi:10.1016/j.epsl.2011.05.006

Boujibar, A., Andrault, D., Bouhifd, M. A., Bolfan-Casanova, N., Devidal, J. L., and Trcera, N. (2014). Metal–silicate partitioning of sulphur, new experimental and thermodynamic constraints on planetary accretion. *Earth Planet. Sci. Lett.* 391, 42–54. doi:10.1016/j.epsl.2014.01.021

Boujibar, A., Driscoll, P., and Fei, Y. (2020). Super-Earth internal structures and initial thermal states. *J. Geophys. Res. Planets* 125 (5), e2019JE006124. doi:10.1029/2019je006124

Boulliung, J., Füri, E., Dalou, C., Tissandier, L., Zimmermann, L., and Marrocchi, Y. (2020). Oxygen fugacity and melt composition controls on nitrogen solubility in silicate melts. *Geochimica Cosmochimica Acta* 284, 120–133. doi:10.1016/j.gca.2020.06.020

Bower, D. J., Hakim, K., Sossi, P. A., and Sanan, P. (2022). Retention of water in terrestrial magma oceans and carbon-rich early atmospheres. *Planet. Sci. J.* 3 (4), 93. doi:10.3847/psj/ac5fb1

Brennan, M. C., Fischer, R. A., and Irving, J. C. E. (2020). Core formation and geophysical properties of Mars. *Earth Planet. Sci. Lett.* 530, 115923. doi:10.1016/j.epsl.2019.115923

Busemann, H., Lorenzetti, S., and Eugster, O. (2006). Noble gases in D'Orbigny, Sahara 99555 and D'Orbigny glass—evidence for early planetary processing on the angrite parent body. *Geochimica cosmochimica acta* 70 (21), 5403–5425. doi:10.1016/j.gca.2006.08.015

Campbell, A. J., Danielson, L., Righter, K., Seagle, C. T., Wang, Y., and Prakapenka, V. B. (2009). High pressure effects on the iron–iron oxide and nickel–nickel oxide oxygen fugacity buffers. *Earth Planet. Sci. Lett.* 286 (3-4), 556–564. doi:10.1016/j.epsl.2009.07.022

Caracas, R., and Stewart, S. T. (2023). No magma ocean surface after giant impacts between rocky planets. *Earth Planet. Sci. Lett.* 608, 118014. doi:10.1016/j.epsl.2023.118014

Carroll, M. R., and Stolper, E. M. (1993). Noble gas solubilities in silicate melts and glasses: new experimental results for argon and the relationship between solubility and ionic porosity. *Geochimica Cosmochimica Acta* 57 (23-24), 5039–5051. doi:10.1016/0016-7037(93)90606-w

Catling, D. C., Zahnle, K. J., and McKay, C. P. (2001). Biogenic methane, hydrogen escape, and the irreversible oxidation of early Earth. *Science* 293 (5531), 839–843. doi:10.1126/science.1061976

Chabot, N. L. (2004). Sulfur contents of the parental metallic cores of magmatic iron meteorites. *Geochimica Cosmochimica Acta* 68 (17), 3607–3618. doi:10.1016/j.gca.2004.03.023

Chabot, N. L., and Agee, C. B. (2003). Core formation in the Earth and Moon: new experimental constraints from V, Cr, and Mn. *Geochimica Cosmochimica Acta* 67 (11), 2077–2091. doi:10.1016/s0016-7037(02)01272-3

Chachan, Y., and Stevenson, D. J. (2018). On the role of dissolved gases in the atmosphere retention of low-mass low-density planets. *Astrophysical J.* 854 (1), 21. doi:10.3847/1538-4357/aaa459

Chambers, J. E., and Wetherill, G. W. (1998). Making the terrestrial planets: N-body integrations of planetary embryos in three dimensions. *Icarus* 136 (2), 304–327. doi:10.1006/icar.1998.6007

Chao, K.-H., deGraffenried, R., Lach, M., Nelson, W., Truax, K., and Gaidos, E. (2021). Lava worlds: from early earth to exoplanets. *Geochemistry* 81 (2), 125735. doi:10.1016/j.chemer.2020.125735

Chi, H., Dasgupta, R., Duncan, M. S., and Shimizu, N. (2014). Partitioning of carbon between Fe-rich alloy melt and silicate melt in a magma ocean–implications for the abundance and origin of volatiles in Earth, Mars, and the Moon. *Geochimica Cosmochimica Acta* 139, 447–471. doi:10.1016/j.gca.2014.04.046

Chidester, B. A., Lock, S. J., Swadba, K. E., Rahman, Z., Righter, K., and Campbell, A. J. (2022). The lithophile element budget of Earth's core. *Geochem. Geophys. Geosystems* 23 (2), e2021GC009986. doi:10.1029/2021gc009986

Clesi, V., Bouhifd, M. A., Bolfan-Casanova, N., Manthilake, G., Schiavi, F., Raepsaet, C., et al. (2018). Low hydrogen contents in the cores of terrestrial planets. *Sci. Adv.* 4 (3), e1701876. doi:10.1126/sciadv.1701876

Corgne, A., Keshav, S., Wood, B. J., McDonough, W. F., and Fei, Y. (2008a). Metal–silicate partitioning and constraints on core composition and oxygen fugacity during Earth accretion. *Geochimica Cosmochimica Acta* 72 (2), 574–589. doi:10.1016/j.gca.2007.10.006

Corgne, A., Siebert, J., and Badro, J. (2009). Oxygen as a light element: a solution to single-stage core formation. *Earth Planet. Sci. Lett.* 288 (1-2), 108–114. doi:10.1016/j.epsl.2009.09.012

Corgne, A., Wood, B. J., and Fei, Y. (2008b). C-and S-rich molten alloy immiscibility and core formation of planetesimals. *Geochimica Cosmochimica Acta* 72 (9), 2409–2416. doi:10.1016/j.gca.2008.03.001

Dahl, T. W., and Stevenson, D. J. (2010). Turbulent mixing of metal and silicate during planet accretion—and interpretation of the Hf–W chronometer. *Earth Planet. Sci. Lett.* 295 (1-2), 177–186. doi:10.1016/j.epsl.2010.03.038

Dalou, C., Deligny, C., and Füri, E. (2022). Nitrogen isotope fractionation during magma ocean degassing: tracing the composition of early Earth's atmosphere. *Geochem. Perspect. Lett.* 20, 27–31. doi:10.7185/geochemlet.2204

Dalou, C., Füri, E., Deligny, C., Piani, L., Caumon, M. C., Laumonier, M., et al. (2019). Redox control on nitrogen isotope fractionation during planetary core formation. *Proc. Natl. Acad. Sci.* 116 (29), 14485–14494. doi:10.1073/pnas.1820719116

Dalou, C., Hirschmann, M. M., von der Handt, A., Mosenfelder, J., and Armstrong, L. S. (2017). Nitrogen and carbon fractionation during core–mantle differentiation at shallow depth. *Earth Planet. Sci. Lett.* 458, 141–151. doi:10.1016/j.epsl.2016.10.026

Dasgupta, R., Chi, H., Shimizu, N., Buono, A. S., and Walker, D. (2013). Carbon solution and partitioning between metallic and silicate melts in a shallow magma ocean: implications for the origin and distribution of terrestrial carbon. *Geochimica Cosmochimica Acta* 102, 191–212. doi:10.1016/j.gca.2012.10.011

Dasgupta, R., and Grewal, D. S. (2019). Origin and early differentiation of carbon and associated life-essential volatile elements on Earth. *Deep carbon* 2019, 4–39. doi:10.1017/9781108677950.002

Dasgupta, R., and Hirschmann, M. M. (2010). The deep carbon cycle and melting in Earth's interior. *Earth Planet. Sci. Lett.* 298 (1-2), 1–13. doi:10.1016/j.epsl.2010.06.039

Davies, E. J., Carter, P. J., Root, S., Kraus, R. G., Spaulding, D. K., Stewart, S. T., et al. (2020). Silicate melting and vaporization during rocky Planet Formation. *J. Geophys. Res. Planets* 125 (2), e2019JE006227. doi:10.1029/2019je006227

Davis, A. H., Solomatova, N. V., Campbell, A. J., and Caracas, R. (2022). The speciation and coordination of a deep earth carbonate-silicate-metal melt. *J. Geophys. Res. Solid Earth* 127 (3), e2021JB023314. doi:10.1029/2021jb023314

Deguen, R., Landeau, M., and Olson, P. (2014). Turbulent metal–silicate mixing, fragmentation, and equilibration in magma oceans. *Earth Planet. Sci. Lett.* 391, 274–287. doi:10.1016/j.epsl.2014.02.007

Deligny, C., Füri, E., and Deloule, E. (2021). Origin and timing of volatile delivery (N, H) to the angrite parent body: constraints from *in situ* analyses of melt inclusions. *Geochimica Cosmochimica Acta* 313, 243–256. doi:10.1016/j.gca.2021.07.038

Ding, F., and Wordsworth, R. D. (2022). Prospects for water vapor detection in the atmospheres of temperate and arid rocky exoplanets around M-dwarf stars. *Astrophysical J. Lett.* 925 (1), L8. doi:10.3847/2041-8213/ac4a5d

Dixon, J. E., and Stolper, E. M. (1995). An experimental study of water and carbon dioxide solubilities in mid-ocean ridge basaltic liquids. Part II: applications to degassing. *J. petrology* 36 (6), 1633–1646.

Dixon, J. E., Stolper, E. M., and Holloway, J. R. (1995). An experimental study of water and carbon dioxide solubilities in mid-ocean ridge basaltic liquids. Part I: calibration and solubility models. *J. Petrology* 36 (6), 1607–1631.

Dorn, C., and Lichtenberg, T. (2021). Hidden water in magma ocean exoplanets. *Astrophysical J. Lett.* 922 (1), L4. doi:10.3847/2041-8213/ac33af

Dreibus, G., and Wanke, H. (1985). Mars, a volatile-rich planet. *Meteoritics* 20, 367–381.

Duncan, M. S., Dasgupta, R., and Tsuno, K. (2017). Experimental determination of CO2 content at graphite saturation along a natural basalt-peridotite melt join: implications for the fate of carbon in terrestrial magma oceans. *Earth Planet. Sci. Lett.* 466, 115–128. doi:10.1016/j.epsl.2017.03.008

Elkins-Tanton, L. T. (2012). Magma oceans in the inner solar system. *Annu. Rev. Earth Planet. Sci.* 40, 113–139. doi:10.1146/annurev-earth-042711-105503

Fei, Y., Prewitt, C. T., Mao, H. k., and Bertka, C. M. (1995). Structure and density of FeS at high pressure and high temperature and the internal structure of Mars. *Science* 268 (5219), 1892–1894. doi:10.1126/science.268.5219.1892

Fichtner, C. E., Schmidt, M. W., Liebske, C., Bouvier, A. S., and Baumgartner, L. P. (2021). Carbon partitioning between metal and silicate melts during Earth accretion. *Earth Planet. Sci. Lett.* 554, 116659. doi:10.1016/j.epsl.2020.116659

Fincham, C. J. B., and Richardson, F. D. (1954). The behaviour of sulphur in silicate and aluminate melts. *Proc. R. Soc. Lond. Ser. A. Math. Phys. Sci.* 223 (1152), 40–62.

Fine, G., and Stolper, E. (1986). Dissolved carbon dioxide in basaltic glasses: concentrations and speciation. *Earth Planet. Sci. Lett.* 76 (3-4), 263–278. doi:10.1016/0012-821x(86)90078-6

Fischer, R. A., Cottrell, E., Hauri, E., Lee, K. K. M., and Le Voyer, M. (2020). The carbon content of Earth and its core. *Proc. Natl. Acad. Sci.* 117 (16), 8743–8749. doi:10.1073/pnas.1919930117

Fischer, R. A., Nakajima, Y., Campbell, A. J., Frost, D. J., Harries, D., Langenhorst, F., et al. (2015). High pressure metal–silicate partitioning of Ni, Co, V, Cr, Si, and O. *Geochimica Cosmochimica Acta* 167, 177–194. doi:10.1016/j.gca.2015.06.026

Fulton, B. J., Petigura, E. A., Howard, A. W., Isaacson, H., Marcy, G. W., Cargile, P. A., et al. (2017). The California-Kepler survey. III. A gap in the radius distribution of small planets. *Astronomical J.* 154 (3), 109. doi:10.3847/1538-3881/aa80eb

Gaffney, J. A., Yang, L., and Ali, S. (2022). Constraining model uncertainty in plasma equation-of-state models with a physics-constrained Gaussian process. arXiv preprint arXiv:2207.00668. Available at: https://doi.org/10.48550/arXiv.2207.00668.

Gaillard, F., Bernadou, F., Roskosz, M., Bouhifd, M. A., Marrocchi, Y., Iacono-Marziano, G., et al. (2022). Redox controls during magma ocean degassing. *Earth Planet. Sci. Lett.* 577, 117255. doi:10.1016/j.epsl.2021.117255

Gaillard, F., Bouhifd, M. A., Füri, E., Malavergne, V., Marrocchi, Y., Noack, L., et al. (2021). The diverse planetary ingassing/outgassing paths produced over billions of years of magmatic activity. *Space Sci. Rev.* 217 (1), 22–54. doi:10.1007/s11214-021-00802-1

Ganguly, J. (2008). *Thermodynamics in earth and planetary sciences*. Heidelberg: Springer.

Ghiorso, M. S., and Gualda, G. A. R. (2015). An H 2 O–CO 2 mixed fluid saturation model compatible with rhyolite-MELTS. *Contributions Mineralogy Petrology* 169, 53–30. doi:10.1007/s00410-015-1141-8

Gounelle, M., and Zolensky, M. E. (2014). The Orgueil meteorite: 150 years of history. *Meteorit. Planet. Sci.* 49 (10), 1769–1794. doi:10.1111/maps.12351

Greene, T. P., Bell, T. J., Ducrot, E., Dyrek, A., Lagage, P. O., and Fortney, J. J. (2023). Thermal emission from the Earth-sized exoplanet TRAPPIST-1 b using JWST. *Nature* 618 (7963), 39–42. doi:10.1038/s41586-023-05951-7

Grewal, D. S. (2022). Origin of nitrogen isotopic variations in the rocky bodies of the solar system. *Astrophysical J.* 937 (2), 123. doi:10.3847/1538-4357/ac8eb4

Grewal, D. S., and Asimow, P. D. (2023). Origin of the superchondritic carbon/nitrogen ratio of the bulk silicate Earth– an outlook from iron meteorites. *Geochimica Cosmochimica Acta* 344, 146–159. doi:10.1016/j.gca.2023.01.012

Grewal, D. S., Dasgupta, R., and Aithala, S. (2021a). The effect of carbon concentration on its core-mantle partitioning behavior in inner Solar System rocky bodies. *Earth Planet. Sci. Lett.* 571, 117090. doi:10.1016/j.epsl.2021.117090

Grewal, D. S., Dasgupta, R., and Farnell, A. (2020). The speciation of carbon, nitrogen, and water in magma oceans and its effect on volatile partitioning between major reservoirs of the Solar System rocky bodies. *Geochimica Cosmochimica Acta* 280, 281–301. doi:10.1016/j.gca.2020.04.023

Grewal, D. S., Dasgupta, R., Holmes, A. K., Costin, G., Li, Y., and Tsuno, K. (2019a). The fate of nitrogen during core-mantle separation on Earth. *Geochimica cosmochimica acta* 251, 87–115. doi:10.1016/j.gca.2019.02.009

Grewal, D. S., Dasgupta, R., Hough, T., and Farnell, A. (2021b). Rates of protoplanetary accretion and differentiation set nitrogen budget of rocky planets. *Nat. Geosci.* 14 (6), 369–376. doi:10.1038/s41561-021-00733-0

Grewal, D. S., Dasgupta, R., and Marty, B. (2021c). A very early origin of isotopically distinct nitrogen in inner Solar System protoplanets. *Nat. Astron.* 5 (4), 356–364. doi:10.1038/s41550-020-01283-y

Grewal, D. S., Dasgupta, R., Sun, C., Tsuno, K., and Costin, G. (2019b). Delivery of carbon, nitrogen, and sulfur to the silicate Earth by a giant impact. *Sci. Adv.* 5 (1), eaau3669. doi:10.1126/sciadv.aau3669

Grewal, D. S., Seales, J. D., and Dasgupta, R. (2022a). Internal or external magma oceans in the earliest protoplanets–Perspectives from nitrogen and carbon fractionation. *Earth Planet. Sci. Lett.* 598, 117847. doi:10.1016/j.epsl.2022.117847

Grewal, D. S., Sun, T., Aithala, S., Hough, T., Dasgupta, R., Yeung, L. Y., et al. (2022b). Limited nitrogen isotopic fractionation during core-mantle differentiation in rocky protoplanets and planets. *Geochimica Cosmochimica Acta* 338, 347–364. doi:10.1016/j.gca.2022.10.025

Gualda, G. A. R., and Ghiorso, M. S. (2015). MELTS _ E xcel: AM icrosoft E xcel-based MELTS interface for research and teaching of magma properties and evolution. *Geochem. Geophys. Geosystems* 16 (1), 315–324. doi:10.1002/2014gc005545

Halliday, A. N. (2013). The origins of volatiles in the terrestrial planets. *Geochimica Cosmochimica Acta* 105, 146–171. doi:10.1016/j.gca.2012.11.015

Helgeson, H. C., and Kirkham, D. H. (1974). Theoretical prediction of the thermodynamic behavior of aqueous electrolytes at high pressures and temperatures; I, Summary of the thermodynamic/electrostatic properties of the solvent. *Am. J. Sci.* 274 (10), 1089–1198. doi:10.2475/ajs.274.10.1089

Hier-Majumder, S., and Hirschmann, M. M. (2017). The origin of volatiles in the E arth's mantle. *Geochem. Geophys. Geosystems* 18 (8), 3078–3092. doi:10.1002/2017gc006937

Hirschmann, M. M. (2012). Magma ocean influence on early atmosphere mass and composition. *Earth Planet. Sci. Lett.* 341, 48–57. doi:10.1016/j.epsl.2012.06.015

Hirschmann, M. M. (2016). Constraints on the early delivery and fractionation of Earth's major volatiles from C/H, C/N, and C/S ratios. *Am. Mineralogist* 101 (3), 540–553. doi:10.2138/am-2016-5452

Hirschmann, M. M. (2018). Comparative deep Earth volatile cycles: the case for C recycling from exosphere/mantle fractionation of major (H2O, C, N) volatiles and from H2O/Ce, CO2/Ba, and CO2/Nb exosphere ratios. *Earth Planet. Sci. Lett.* 502, 262–273. doi:10.1016/j.epsl.2018.08.023

Huang, D., and Badro, J. (2018). Fe-Ni ideality during core formation on Earth. *Am. Mineralogist* 103 (10), 1707–1710. doi:10.2138/am-2018-6651

Iacono-Marziano, G., Morizet, Y., Le Trong, E., and Gaillard, F. (2012). New experimental data and semi-empirical parameterization of H2O–CO2 solubility in mafic melts. *Geochimica Cosmochimica Acta* 97, 1–23. doi:10.1016/j.gca.2012.08.035

Iacono-Marziano, G., Paonita, A., Rizzo, A., Scaillet, B., and Gaillard, F. (2010). Noble gas solubilities in silicate melts: new experimental results and a comprehensive model of the effects of liquid composition, temperature and pressure. *Chem. Geol.* 279 (3-4), 145–157. doi:10.1016/j.chemgeo.2010.10.017

Iacovino, K., Matthews, S., Wieser, P. E., Moore, G. M., and Bégué, F. (2021). VESIcal Part I: an open-source thermodynamic model engine for mixed volatile (H2O-CO2) solubility in silicate melts. *Earth Space Sci.* 8 (11), e2020EA001584. doi:10.1029/2020ea001584

Ih, J., Kempton, E. M. R., Whittaker, E. A., and Lessard, M. (2023). Constraining the thickness of TRAPPIST-1 b's atmosphere from its JWST secondary eclipse observation at 15 μm. *Astrophysical J. Lett.* 952 (1), L4. doi:10.3847/2041-8213/ace03b

Ikoma, M., Elkins-Tanton, L., Hamano, K., and Suckale, J. (2018). Water partitioning in planetary embryos and protoplanets with magma oceans. *Space Sci. Rev.* 214, 76–28. doi:10.1007/s11214-018-0508-3

Jackson, C. R., Bennett, N. R., Du, Z., Cottrell, E., and Fei, Y. (2018). Early episodes of high-pressure core formation preserved in plume mantle. *Nature* 553, 491.

Jackson, C. R. M., Cottrell, E., Du, Z., Bennett, N., and Fei, Y. (2021). High pressure redistribution of nitrogen and sulfur during planetary stratification. *Geochem. Perspect. Lett.* 18, 37–42. doi:10.7185/geochemlet.2122

Jambon, A., Weber, H., and Braun, O. (1986). Solubility of He, Ne, Ar, Kr and Xe in a basalt melt in the range 1250–1600 C. Geochemical implications. *Geochimica Cosmochimica Acta* 50 (3), 401–408. doi:10.1016/0016-7037(86)90193-6

Jana, D., and Walker, D. (1997). The influence of sulfur on partitioning of siderophile elements. *Geochimica Cosmochimica Acta* 61 (24), 5255–5277. doi:10.1016/s0016-7037(97)00307-4

Johansen, A., Ronnet, T., Bizzarro, M., Schiller, M., Lambrechts, M., Nordlund, Å., et al. (2021). A pebble accretion model for the formation of the terrestrial planets in the Solar System. *Sci. Adv.* 7 (8), eabc0444. doi:10.1126/sciadv.abc0444

Jones, J. H., and Drake, M. J. (1986). Geochemical constraints on core formation in the Earth. *Nature* 322 (6076), 221–228. doi:10.1038/322221a0

Kaltenegger, L., Pepper, J., Stassun, K., and Oelkers, R. (2019). TESS habitable zone star catalog. *Astrophysical J. Lett.* 874 (1), L8. doi:10.3847/2041-8213/ab0e8d

Karato, S.-I., and Murthy, V. R. (1997). Core formation and chemical equilibrium in the Earth—I. Physical considerations. *Phys. Earth Planet. Interiors* 100 (1-4), 61–79. doi:10.1016/s0031-9201(96)03232-3

Kempton, E. M. R., Lessard, M., Malik, M., Rogers, L. A., Futrowsky, K. E., Ih, J., et al. (2023). Where are the water worlds? self-consistent models of water-rich exoplanet atmospheres. *Astrophysical J.* 953 (1), 57. doi:10.3847/1538-4357/ace10d

Keppler, H., and Golabek, G. (2019). Graphite floatation on a magma ocean and the fate of carbon during core formation. *Geochem. Perspect. Lett.* 11, 12–17. doi:10.7185/geochemlet.1918

Kimura, T., and Ikoma, M. (2022). Predicted diversity in water content of terrestrial exoplanets orbiting M dwarfs. *Nat. Astron.* 6 (11), 1296–1307. doi:10.1038/s41550-022-01781-1

Kite, E. S., Fegley Jr., B., Schaefer, L., and Ford, E. B. (2020). Atmosphere origins for exoplanet sub-neptunes. *Astrophysical J.* 891 (2), 111. doi:10.3847/1538-4357/ab6ffb

Kite, E. S., and Schaefer, L. (2021). Water on hot rocky exoplanets. *Astrophysical J. Lett.* 909 (2), L22. doi:10.3847/2041-8213/abe7dc

Konschak, A., and Keppler, H. (2014). The speciation of carbon dioxide in silicate melts. *Contributions Mineralogy Petrology* 167, 998–1013. doi:10.1007/s00410-014-0998-2

Kovačević, T., González-Cataldo, F., Stewart, S. T., and Militzer, B. (2022). Miscibility of rock and ice in the interiors of water worlds. *Sci. Rep.* 12 (1), 13055. doi:10.1038/s41598-022-16816-w

Kuwahara, H., Itoh, S., Nakada, R., and Irifune, T. (2019). The effects of carbon concentration and silicate composition on the metal-silicate partitioning of carbon in a shallow magma ocean. *Geophys. Res. Lett.* 46 (16), 9422–9429. doi:10.1029/2019gl084254

Kuwahara, H., Itoh, S., Suzumura, A., Nakada, R., and Irifune, T. (2021). Nearly carbon-saturated magma oceans in planetary embryos during core formation. *Geophys. Res. Lett.* 48 (10), e2021GL092389. doi:10.1029/2021gl092389

Lesne, P., Kohn, S. C., Blundy, J., Witham, F., Botcharnikov, R. E., and Behrens, H. (2011). Experimental simulation of closed-system degassing in the system basalt–H2O–CO2–S–Cl. *J. Petrology* 52 (9), 1737–1762. doi:10.1093/petrology/egr027

Lewis, J. A., Jones, R. H., and Brearley, A. J. (2022). Plagioclase alteration and equilibration in ordinary chondrites: metasomatism during thermal metamorphism. *Geochimica Cosmochimica Acta* 316, 201–229. doi:10.1016/j.gca.2021.10.004

Li, J., and Agee, C. B. (1996). Geochemistry of mantle–core differentiation at high pressure. *Nature* 381 (6584), 686–689. doi:10.1038/381686a0

Li, J., and Agee, C. B. (2001). Element partitioning constraints on the light element composition of the Earth's core. *Geophys. Res. Lett.* 28 (1), 81–84. doi:10.1029/2000gl012114

Li, J., Bergin, E. A., Blake, G. A., Ciesla, F. J., and Hirschmann, M. M. (2021). Earth's carbon deficit caused by early loss through irreversible sublimation. *Sci. Adv.* 7 (14), eabd3632. doi:10.1126/sciadv.abd3632

Li, Y., Dasgupta, R., and Tsuno, K. (2015). The effects of sulfur, silicon, water, and oxygen fugacity on carbon solubility and partitioning in Fe-rich alloy and silicate melt systems at 3 GPa and 1600 °C: implications for core–mantle differentiation and degassing of magma oceans and reduced planetary mantles. *Earth Planet. Sci. Lett.* 415, 54–66. doi:10.1016/j.epsl.2015.01.017

Li, Y., Dasgupta, R., Tsuno, K., Monteleone, B., and Shimizu, N. (2016b). Carbon and sulfur budget of the silicate Earth explained by accretion of differentiated planetary embryos. *Nat. Geosci.* 9 (10), 781–785. doi:10.1038/ngeo2801

Li, Y., Vočadlo, L., Sun, T., and Brodholt, J. P. (2020). The Earth's core as a reservoir of water. *Nat. Geosci.* 13 (6), 453–458. doi:10.1038/s41561-020-0578-1

Li, Y., Wiedenbeck, M., Shcheka, S., and Keppler, H. (2013). Nitrogen solubility in upper mantle minerals. *Earth Planet. Sci. Lett.* 377-378, 311–323. doi:10.1016/j.epsl.2013.07.013

Li, Y.-F., Marty, B., Shcheka, S., Zimmermann, L., and Keppler, H. (2016a). Nitrogen isotope fractionation during terrestrial core-mantle separation. *Geochem. Perspect. Lett.* 2, 138–147. doi:10.7185/geochemlet.1614

Libourel, G., Marty, B., and Humbert, F. (2003). Nitrogen solubility in basaltic melt. Part I. Effect of oxygen fugacity. *Geochimica Cosmochimica Acta* 67 (21), 4123–4135. doi:10.1016/s0016-7037(03)00259-x

Lichtenberg, T. (2022). Redox hysteresis of super-Earth exoplanets from magma ocean circulation. *Astrophysical J. Lett.* 914 (1), L4. doi:10.3847/2041-8213/ac0146

Lichtenberg, T., Bower, D. J., Hammond, M., Boukrouche, R., Sanan, P., Tsai, S., et al. (2021a). Vertically resolved magma ocean–protoatmosphere evolution: H2, H2O, CO2, CH4, CO, O2, and N2 as primary absorbers. *J. Geophys. Res. Planets* 126 (2), e2020JE006711. doi:10.1029/2020je006711

Lichtenberg, T., Drażkowska, J., Schönbächler, M., Golabek, G. J., and Hands, T. O. (2021b). Bifurcation of planetary building blocks during Solar System formation. *Science* 371 (6527), 365–370. doi:10.1126/science.abb3091

Lichtenberg, T., Golabek, G. J., Burn, R., Meyer, M. R., Alibert, Y., Gerya, T. V., et al. (2019). A water budget dichotomy of rocky protoplanets from 26Al-heating. *Nat. Astron.* 3 (4), 307–313. doi:10.1038/s41550-018-0688-5

Lichtenberg, T., and Krijt, S. (2021). System-level fractionation of carbon from disk and planetesimal processing. *Astrophysical J. Lett.* 913 (2), L20. doi:10.3847/2041-8213/abfdce

Lichtenberg, T., Schaefer, L. K., Nakajima, M., and Fischer, R. A. (2022). Geophysical evolution during rocky Planet Formation. arXiv preprint arXiv:2203.10023. Available at: https://doi.org/10.48550/arXiv.2203.10023.

Luth, R. W., O Mysen, B., and Virgo, D. (1987). Raman spectroscopic study of the solubility behavior of H2 in the system Na2O-Al2O3-SiO2-H2. *Am. Mineralogist* 72 (5-6), 481–486.

Lv, C., and Liu, J. (2022). Early planetary processes and light elements in iron-dominated cores. *Acta Geochim.* 41 (4), 625–649. doi:10.1007/s11631-021-00522-x

Mahan, B., Siebert, J., Blanchard, I., Badro, J., Kubik, E., Sossi, P., et al. (2018). Investigating earth's formation history through copper and sulfur metal-silicate partitioning during core-mantle differentiation. *J. Geophys. Res. Solid Earth* 123 (10), 8349–8363. doi:10.1029/2018jb015991

Malavergne, V., Bureau, H., Raepsaet, C., Gaillard, F., Poncet, M., Surblé, S., et al. (2019). Experimental constraints on the fate of H and C during planetary core-mantle differentiation. Implications for the Earth. *Icarus* 321, 473–485. doi:10.1016/j.icarus.2018.11.027

Mann, U., Frost, D. J., and Rubie, D. C. (2009). Evidence for high-pressure core-mantle differentiation from the metal–silicate partitioning of lithophile and weakly-siderophile elements. *Geochimica Cosmochimica Acta* 73 (24), 7360–7386. doi:10.1016/j.gca.2009.08.006

Mansfield, M., Kite, E. S., Hu, R., Koll, D. D. B., Malik, M., Bean, J. L., et al. (2019). Identifying atmospheres on rocky exoplanets through inferred high albedo. *Astrophysical J.* 886 (2), 141. doi:10.3847/1538-4357/ab4c90

Marty, B. (2012). The origins and concentrations of water, carbon, nitrogen and noble gases on Earth. *Earth Planet. Sci. Lett.* 313, 56–66. doi:10.1016/j.epsl.2011.10.040

Matsui, T., and Abe, Y. (1986). Evolution of an impact-induced atmosphere and magma ocean on the accreting Earth. *Nature* 319 (6051), 303–305. doi:10.1038/319303a0

Mavrogenes, J. A., and O'Neill, H.St C. (1999). The relative effects of pressure, temperature and oxygen fugacity on the solubility of sulfide in mafic magmas. *Geochimica Cosmochimica Acta* 63 (7-8), 1173–1180. doi:10.1016/s0016-7037(98)00289-0

McCubbin, F. M., and Barnes, J. J. (2019). Origin and abundances of H2O in the terrestrial planets, Moon, and asteroids. *Earth Planet. Sci. Lett.* 526, 115771. doi:10.1016/j.epsl.2019.115771

McDonough, W. F., and Sun, S. S. (1995). The composition of the Earth. *Chem. Geol.* 120 (3-4), 223–253. doi:10.1016/0009-2541(94)00140-4

Mel'nik, Y. P. (1972). Thermodynamic parameters of compressed gases and metamorphic reaction involving water and carbon dioxide. *Geochim. Inter.* 9, 419–425.

Mikhail, S., and Sverjensky, D. A. (2014). Nitrogen speciation in upper mantle fluids and the origin of Earth's nitrogen-rich atmosphere. *Nat. Geosci.* 7 (11), 816–819. doi:10.1038/ngeo2271

Misener, W., and Schlichting, H. E. (2021). To cool is to keep: residual H/He atmospheres of super-Earths and sub-Neptunes. *Mon. Notices R. Astronomical Soc.* 503 (4), 5658–5674. doi:10.1093/mnras/stab895

Moore, K., and Cowan, N. B. (2020). Keeping M-Earths habitable in the face of atmospheric loss by sequestering water in the mantle. *Mon. Notices R. Astronomical Soc.* 496 (3), 3786–3795. doi:10.1093/mnras/staa1796

Moran, S. E., Stevenson, K. B., Sing, D. K., MacDonald, R. J., Kirk, J., Lustig-Yaeger, J., et al. (2023). High tide or riptide on the cosmic shoreline? A water-rich atmosphere or stellar contamination for the warm super-earth gj 486b from JWST observations. *Astrophysical J. Lett.* 948 (1), L11. doi:10.3847/2041-8213/accb9c

Morard, G., Siebert, J., Andrault, D., Guignot, N., Garbarino, G., Guyot, F., et al. (2013). The Earth's core composition from high pressure density measurements of liquid iron alloys. *Earth Planet. Sci. Lett.* 373, 169–178. doi:10.1016/j.epsl.2013.04.040

Mosenfelder, J. L., Von Der Handt, A., Füri, E., Dalou, C., Hervig, R. L., Rossman, G. R., et al. (2019). Nitrogen incorporation in silicates and metals: results from SIMS, EPMA, FTIR, and laser-extraction mass spectrometry. *Am. Mineralogist J. Earth Planet. Mater.* 104 (1), 31–46. doi:10.2138/am-2019-6533

Mysen, B. O., and Boettcher, A. L. (1975). Melting of a hydrous mantle: I. Phase relations of natural peridotite at high pressures and temperatures with controlled activities of water, carbon dioxide, and hydrogen. *J. Petrology* 16 (1), 520–548. doi:10.1093/petrology/16.1.520

Mysen, B. O., and Virgo, D. (1986). Volatiles in silicate melts at high pressure and temperature: 1. Interaction between OH groups and Si4+, Al3+, Ca2+, Na+ and H+. *Chem. Geol.* 57 (3-4), 303–331. doi:10.1016/0009-2541(86)90056-2

Mysen, B. O., and Yamashita, S. (2010). Speciation of reduced C–O–H volatiles in coexisting fluids and silicate melts determined *in-situ* to ~1.4GPa and 800°C. *Geochimica Cosmochimica Acta* 74 (15), 4577–4588. doi:10.1016/j.gca.2010.05.004

Namur, O., Charlier, B., Holtz, F., Cartier, C., and McCammon, C. (2016). Sulfur solubility in reduced mafic silicate melts: implications for the speciation and distribution of sulfur on Mercury. *Earth Planet. Sci. Lett.* 448, 102–114. doi:10.1016/j.epsl.2016.05.024

Newman, S., and Lowenstern, J. B. (2002). VolatileCalc: a silicate melt–H2O–CO2 solution model written in Visual Basic for excel. *Comput. Geosciences* 28 (5), 597–604. doi:10.1016/s0098-3004(01)00081-4

Nikolaou, A., Katyal, N., Tosi, N., Godolt, M., Grenfell, J. L., and Rauer, H. (2019). What factors affect the duration and outgassing of the terrestrial magma ocean? *Astrophysical J.* 875 (1), 11. doi:10.3847/1538-4357/ab08ed

Nimmo, F., and Schubert, G. (2015). *Thermal and compositional evolution of the core*. Elsevier Amsterdam: Core Dynamics, Treatise on Geophysics, 217–241.

Nowak, M., and Behrens, H. (1995). The speciation of water in haplogranitic glasses and melts determined by *in situ* near-infrared spectroscopy. *Geochimica Cosmochimica Acta* 59 (16), 3445–3450. doi:10.1016/0016-7037(95)00237-t

O'Brien, D. P., Walsh, K. J., Morbidelli, A., Raymond, S. N., and Mandell, A. M. (2014). Water delivery and giant impacts in the 'Grand Tack'scenario. *Icarus* 239, 74–84. doi:10.1016/j.icarus.2014.05.009

Ohtani, E., Yurimoto, H., and Seto, S. (1997). Element partitioning between metallic liquid, silicate liquid, and lower-mantle minerals: implications for core formation of the Earth. *Phys. Earth Planet. interiors* 100 (1-4), 97–114. doi:10.1016/s0031-9201(96)03234-7

Okuchi, T. (1997). Hydrogen partitioning into molten iron at high pressure: implications for Earth's core. *Science* 278 (5344), 1781–1784. doi:10.1126/science.278.5344.1781

O'Neill, H. S. C., and Mavrogenes, J. A. (2002). The sulfide capacity and the sulfur content at sulfide saturation of silicate melts at 1400degreesC and 1 bar. *J. Petrology* 43 (6), 1049–1087. doi:10.1093/petrology/43.6.1049

Otegi, J. F., Bouchy, F., and Helled, R., 2020, Revisited mass-radius relations for exoplanets below 120 M⊕. *Astronomy Astrophysics* 634:A43, doi:10.1051/0004-6361/201936482

Pahlevan, K., Schaefer, L., Elkins-Tanton, L. T., Desch, S. J., and Buseck, P. R. (2022). A primordial atmospheric origin of hydrospheric deuterium enrichment on Mars. *Earth Planet. Sci. Lett.* 595, 117772. doi:10.1016/j.epsl.2022.117772

Papale, P., Moretti, R., and Barbato, D. (2006). The compositional dependence of the saturation surface of H2O+ CO2 fluids in silicate melts. *Chem. Geol.* 229 (1-3), 78–95. doi:10.1016/j.chemgeo.2006.01.013

Peslier, A. H. (2010). A review of water contents of nominally anhydrous natural minerals in the mantles of Earth, Mars and the Moon. *J. Volcanol. Geotherm. Res.* 197 (1-4), 239–258. doi:10.1016/j.jvolgeores.2009.10.006

Peterson, L. D., Newcombe, M. E., Alexander, C. M. O., Wang, J., Sarafian, A. R., Bischoff, A., et al. (2023). The H2O content of the ureilite parent body. *Geochimica Cosmochimica Acta* 340, 141–157. doi:10.1016/j.gca.2022.10.036

Piette, A. A. A., Gao, P., Brugman, K., and Shahar, A. (2023). Rocky planet or water world? Observability of low-density lava world atmospheres. arXiv preprint arXiv: 2306.10100. Available at: https://doi.org/10.48550/arXiv.2306.10100.

Righter, K., and Drake, M. J. (1996). Core formation in earth's moon, Mars, and vesta. *Icarus* 124 (2), 513–529. doi:10.1006/icar.1996.0227

Righter, K., and Drake, M. J. (1997). A magma ocean on Vesta: core formation and petrogenesis of eucrites and diogenites. *Meteorit. Planet. Sci.* 32 (6), 929–944. doi:10.1111/j.1945-5100.1997.tb01582.x

Ringwood, A. E. (1977). Composition of the core and implications for origin of the Earth. *Geochem. J.* 11 (3), 111–135. doi:10.2343/geochemj.11.111

Rogers, J. G., and Owen, J. E. (2021). Unveiling the planet population at birth. *Mon. Notices R. Astronomical Soc.* 503 (1), 1526–1542. doi:10.1093/mnras/stab529

Rose-Weston, L., Brenan, J. M., Fei, Y., Secco, R. A., and Frost, D. J. (2009). Effect of pressure, temperature, and oxygen fugacity on the metal-silicate partitioning of Te, Se, and S: implications for earth differentiation. *Geochimica Cosmochimica Acta* 73 (15), 4598–4615. doi:10.1016/j.gca.2009.04.028

Roskosz, M., Bouhifd, M., Jephcoat, A., Marty, B., and Mysen, B. (2013). Nitrogen solubility in molten metal and silicate at high pressure and temperature. *Geochimica Cosmochimica Acta* 121, 15–28. doi:10.1016/j.gca.2013.07.007

Rubie, D. C., Nimmo, F., and Melosh, H. J. (2007). *Treatise on geophysics. Formation of Earth's core 9*, 51–90.

Rubie, D. C., Frost, D. J., Mann, U., Asahara, Y., Nimmo, F., Tsuno, K., et al. (2011). Heterogeneous accretion, composition and core–mantle differentiation of the Earth. *Earth Planet. Sci. Lett.* 301 (1-2), 31–42. doi:10.1016/j.epsl.2010.11.030

Rubie, D. C., Laurenz, V., Jacobson, S. A., Morbidelli, A., Palme, H., Vogel, A. K., et al. (2016). Highly siderophile elements were stripped from Earth's mantle by iron sulfide segregation. *Science* 353 (6304), 1141–1144. doi:10.1126/science.aaf6919

Rubie, D. C., Melosh, H., Reid, J., Liebske, C., and Righter, K. (2003). Mechanisms of metal–silicate equilibration in the terrestrial magma ocean. *Earth Planet. Sci. Lett.* 205 (3-4), 239–255. doi:10.1016/s0012-821x(02)01044-0

Rudge, J. F., Kleine, T., and Bourdon, B. (2010). Broad bounds on Earth's accretion and core formation constrained by geochemical models. *Nat. Geosci.* 3 (6), 439–443. doi:10.1038/ngeo872

Salvador, A., Massol, H., Davaille, A., Marcq, E., Sarda, P., and Chassefière, E. (2017). The relative influence of H2O and CO2 on the primitive surface conditions and evolution of rocky planets. *J. Geophys. Res. Planets* 122 (7), 1458–1486. doi:10.1002/2017je005286

Salvador, A., and Samuel, H. (2023). Convective outgassing efficiency in planetary magma oceans: insights from computational fluid dynamics. *Icarus* 390, 115265. doi:10.1016/j.icarus.2022.115265

Sasaki, S. (1990). The primary solar-type atmosphere surrounding the accreting Earth: H 2 O-induced high surface temperature. *Orig. Earth* 1990, 195–209.

Schlichting, H. E., and Young, E. D. (2022). Chemical equilibrium between cores, mantles, and atmospheres of super-earths and sub-neptunes and implications for their compositions, interiors, and evolution. *Planet. Sci. J.* 3 (5), 127. doi:10.3847/psj/ac68e6

Schönbächler, M., Carlson, R. W., Horan, M. F., Mock, T. D., and Hauri, E. H. (2010). Heterogeneous accretion and the moderately volatile element budget of earth. *Science* 328 (5980), 884–887. doi:10.1126/science.1186239

Sharp, Z. D. (2017). Nebular ingassing as a source of volatiles to the Terrestrial planets. *Chem. Geol.* 448, 137–150. doi:10.1016/j.chemgeo.2016.11.018

Shi, L., Lu, W., Kagoshima, T., Sano, Y., Gao, Z., Du, Z., et al. (2022). Nitrogen isotope evidence for Earth's heterogeneous accretion of volatiles. *Nat. Commun.* 13 (1), 4769. doi:10.1038/s41467-022-32516-5

Siebert, J., Badro, J., Antonangeli, D., and Ryerson, F. J. (2012). Metal–silicate partitioning of Ni and Co in a deep magma ocean. *Earth Planet. Sci. Lett.* 321-322, 189–197. doi:10.1016/j.epsl.2012.01.013

Solomatov, V. (2007). "Magma Oceans and primordial mantle differentiation," in *Treatise on geophysics: evolution of the Earth*. Editors G. Schubert and D. Stevenson (United States: Elsevier Ltd).

Solomatova, N. V., and Caracas, R. (2019). Pressure-induced coordination changes in a pyrolitic silicate melt from *ab initio* molecular dynamics simulations. *J. Geophys. Res. Solid Earth* 124 (11), 11232–11250. doi:10.1029/2019jb018238

Sossi, P. A., Tollan, P. M., Badro, J., and Bower, D. J. (2023). Solubility of water in peridotite liquids and the prevalence of steam atmospheres on rocky planets. *Earth Planet. Sci. Lett.* 601, 117894. doi:10.1016/j.epsl.2022.117894

Speelmanns, I. M., Schmidt, M. W., and Liebske, C. (2018). Nitrogen solubility in core materials. *Geophys. Res. Lett.* 45 (15), 7434–7443. doi:10.1029/2018gl079130

Stanley, B. D., Hirschmann, M. M., and Withers, A. C. (2014). Solubility of COH volatiles in graphite-saturated martian basalts. *Geochimica Cosmochimica Acta* 129, 54–76. doi:10.1016/j.gca.2013.12.013

Steenstra, E. S., Berndt, J., Klemme, S., Rohrbach, A., Bullock, E., and van Westrenen, W. (2020). An experimental assessment of the potential of sulfide saturation of the source regions of eucrites and angrites: implications for asteroidal models of core formation, late accretion and volatile element depletions. *Geochimica Cosmochimica Acta* 269, 39–62. doi:10.1016/j.gca.2019.10.006

Steenstra, E. S., Knibbe, J., and van Westrenen, W. (2016). Constraints on core formation in Vesta from metal–silicate partitioning of siderophile elements. *Geochimica Cosmochimica Acta* 177, 48–61. doi:10.1016/j.gca.2016.01.002

Steenstra, E. S., and van Westrenen, W. (2020). Geochemical constraints on core-mantle differentiation in Mercury and the aubrite parent body. *Icarus* 340, 113621. doi:10.1016/j.icarus.2020.113621

Stevenson, D. J. (1981). Models of the Earth's core. *Science* 214 (4521), 611–619. doi:10.1126/science.214.4521.611

Stevenson, D. J. (1988). Fluid dynamics of core formation. *Top. Conf. Orig. Earth* 681, 87.

Stolper, E. (1982). The speciation of water in silicate melts. *Geochimica Cosmochimica Acta* 46 (12), 2609–2620. doi:10.1016/0016-7037(82)90381-7

Suer, T.-A., Siebert, J., Remusat, L., Day, J. M. D., Borensztajn, S., Doisneau, B., et al. (2021). Reconciling metal–silicate partitioning and late accretion in the Earth. *Nat. Commun.* 12 (1), 2913–3010. doi:10.1038/s41467-021-23137-5

Suer, T.-A., Siebert, J., Remusat, L., Menguy, N., and Fiquet, G. (2017). A sulfur-poor terrestrial core inferred from metal–silicate partitioning experiments. *Earth Planet. Sci. Lett.* 469, 84–97. doi:10.1016/j.epsl.2017.04.016

Tagawa, S., Sakamoto, N., Hirose, K., Yokoo, S., Hernlund, J., Ohishi, Y., et al. (2021). Experimental evidence for hydrogen incorporation into Earth's core. *Nat. Commun.* 12 (1), 2588. doi:10.1038/s41467-021-22035-0

Tian, F., Toon, O. B., Pavlov, A. A., and De Sterck, H. (2005). A hydrogen-rich early Earth atmosphere. *Science* 308 (5724), 1014–1017. doi:10.1126/science.1106983

Tonks, W. B., and Melosh, H. J. (1993). Magma ocean formation due to giant impacts. *J. Geophys. Res. Planets* 98 (E3), 5319–5333. doi:10.1029/92je02726

Tsuno, K., Grewal, D. S., and Dasgupta, R. (2018). Core-mantle fractionation of carbon in Earth and Mars: the effects of sulfur. *Geochimica Cosmochimica Acta* 238, 477–495. doi:10.1016/j.gca.2018.07.010

Vander Kaaden, K. E., and McCubbin, F. M. (2015). Exotic crust formation on Mercury: consequences of a shallow, FeO-poor mantle. *J. Geophys. Res. Planets* 120 (2), 195–209. doi:10.1002/2014je004733

Vazan, A., Sari, R., and Kessel, R. (2022). A new perspective on the interiors of ice-rich planets: ice–rock mixture instead of ice on top of rock. *Astrophysical J.* 926 (2), 150. doi:10.3847/1538-4357/ac458c

Wade, J., and Wood, B. J. (2001). The Earth's 'missing'niobium may be in the core. *Nature* 409 (6816), 75–78. doi:10.1038/35051064

Wade, J., and Wood, B. J. (2005). Core formation and the oxidation state of the Earth. *Earth Planet. Sci. Lett.* 236 (1-2), 78–95. doi:10.1016/j.epsl.2005.05.017

Wade, J., Wood, B. J., and Tuff, J. (2012). Metal–silicate partitioning of Mo and W at high pressures and temperatures: evidence for late accretion of sulphur to the Earth. *Geochimica Cosmochimica Acta* 85, 58–74. doi:10.1016/j.gca.2012.01.010

Wahl, S. M., and Militzer, B. (2015). High-temperature miscibility of iron and rock during terrestrial planet formation. *Earth Planet. Sci. Lett.* 410, 25–33. doi:10.1016/j.epsl.2014.11.014

Walter, M. J., and Cottrell, E. (2013). Assessing uncertainty in geochemical models for core formation in Earth. *Earth Planet. Sci. Lett.* 365, 165–176. doi:10.1016/j.epsl.2013.01.014

Wang, Z., and Becker, H. (2013). Ratios of S, Se and Te in the silicate Earth require a volatile-rich late veneer. *Nature* 499 (7458), 328–331. doi:10.1038/nature12285

Wetzel, D. T., Rutherford, M. J., Jacobsen, S. D., Hauri, E. H., and Saal, A. E. (2013). Degassing of reduced carbon from planetary basalts. *Proc. Natl. Acad. Sci.* 110 (20), 8010–8013. doi:10.1073/pnas.1219266110

Wolf, A. S., Jäggi, N., Sossi, P. A., and Bower, D. J. (2022). VapoRock: thermodynamics of vaporized silicate melts for modeling volcanic outgassing and magma ocean atmospheres. arXiv preprint arXiv:2208.09582. Available at: https://doi.org/10.48550/arXiv.2208.09582.

Wood, B. J. (2008). Accretion and core formation: constraints from metal–silicate partitioning. *Philosophical Trans. R. Soc. A Math. Phys. Eng. Sci.* 366 (1883), 4339–4355. doi:10.1098/rsta.2008.0115

Wood, B. J., Walter, M. J., and Wade, J. (2006). Accretion of the Earth and segregation of its core. *Nature* 441 (7095), 825–833. doi:10.1038/nature04763

Wordsworth, R., and Kreidberg, L. (2022). Atmospheres of rocky exoplanets. *Annu. Rev. Astronomy Astrophysics* 60, 159–201. doi:10.1146/annurev-astro-052920-125632

Wordsworth, R. D. (2016). Atmospheric nitrogen evolution on earth and Venus. *Earth Planet. Sci. Lett.* 447, 103–111. doi:10.1016/j.epsl.2016.04.002

Wu, J., Desch, S. J., Schaefer, L., Elkins-Tanton, L. T., Pahlevan, K., and Buseck, P. R. (2018). Origin of Earth's water: chondritic inheritance plus nebular ingassing and storage of hydrogen in the core. *J. Geophys. Res. Planets* 123 (10), 2691–2712. doi:10.1029/2018je005698

Xue, X., and Kanzaki, M. (2004). Dissolution mechanisms of water in depolymerized silicate melts: constraints from 1H and 29Si NMR spectroscopy and *ab initio* calculations. *Geochimica Cosmochimica Acta* 68 (24), 5027–5057. doi:10.1016/j.gca.2004.08.016

Yoshioka, T., Nakashima, D., Nakamura, T., Shcheka, S., and Keppler, H. (2019). Carbon solubility in silicate melts in equilibrium with a CO-CO2 gas phase and graphite. *Geochimica cosmochimica acta* 259, 129–143. doi:10.1016/j.gca.2019.06.007

Young, E. D., Shahar, A., and Schlichting, H. E. (2023). Earth shaped by primordial H2 atmospheres. *Nature* 616 (7956), 306–311. doi:10.1038/s41586-023-05823-0

Yuan, L., and Steinle-Neumann, G. (2020). Strong sequestration of hydrogen into the Earth's core during planetary differentiation. *Geophys. Res. Lett.* 47 (15), e2020GL088303. doi:10.1029/2020gl088303

Zahnle, K., Schaefer, L., and Fegley, B. (2010). Earth's earliest atmospheres. *Cold Spring Harb. Perspect. Biol.* 2 (10), a004895. doi:10.1101/cshperspect.a004895

Zahnle, K. J., Kasting, J. F., and Pollack, J. B. (1988). Evolution of a steam atmosphere during earth's accretion. *Icarus* 74 (1), 62–97. doi:10.1016/0019-1035(88)90031-0

Zahnle, K. J., Lupu, R., Catling, D. C., and Wogan, N. (2020). Creation and evolution of impact-generated reduced atmospheres of early Earth. *Planet. Sci. J.* 1 (1), 11. doi:10.3847/psj/ab7e2c

Zhang, H. L., Cottrell, E., Solheid, P. A., Kelley, K. A., and Hirschmann, M. M. (2018). Determination of Fe3+/ΣFe of XANES basaltic glass standards by Mössbauer spectroscopy and its application to the oxidation state of iron in MORB. *Chem. Geol.* 479, 166–175. doi:10.1016/j.chemgeo.2018.01.006

Zhang, M., and Li, Y. (2021). Breaking of Henry's law for sulfide liquid–basaltic melt partitioning of Pt and Pd. *Nat. Commun.* 12 (1), 5994. doi:10.1038/s41467-021-26311-x

Zieba, S., Kreidberg, L., Ducrot, E., Gillon, M., Morley, C., Schaefer, L., et al. (2023). No thick carbon dioxide atmosphere on the rocky exoplanet TRAPPIST-1 c. *Nature* 620, 746–749. doi:10.1038/s41586-023-06232-z